\RequirePackage{fix-cm}
\documentclass[twocolumn]{svjour3}          
\smartqed  
\usepackage{graphicx}
\usepackage{soul}
\usepackage{color}
\usepackage{listings}
\usepackage{subfig}
\usepackage[T1]{fontenc}
\usepackage[table,xcdraw]{xcolor}
\usepackage{breakcites}
\usepackage[colorlinks, linktocpage=true, linkcolor=blue, citecolor=blue, filecolor=black,urlcolor=blue]{hyperref}
\usepackage[round]{natbib}
\usepackage{etoolbox} 

\makeatletter

\patchcmd{\NAT@citex}
  {\@citea\NAT@hyper@{%
     \NAT@nmfmt{\NAT@nm}%
     \hyper@natlinkbreak{\NAT@aysep\NAT@spacechar}{\@citeb\@extra@b@citeb}%
     \NAT@date}}
  {\@citea\NAT@nmfmt{\NAT@nm}%
   \NAT@aysep\NAT@spacechar\NAT@hyper@{\NAT@date}}{}{}

\patchcmd{\NAT@citex}
  {\@citea\NAT@hyper@{%
     \NAT@nmfmt{\NAT@nm}%
     \hyper@natlinkbreak{\NAT@spacechar\NAT@@open\if*#1*\else#1\NAT@spacechar\fi}%
       {\@citeb\@extra@b@citeb}%
     \NAT@date}}
  {\@citea\NAT@nmfmt{\NAT@nm}%
   \NAT@spacechar\NAT@@open\if*#1*\else#1\NAT@spacechar\fi\NAT@hyper@{\NAT@date}}
  {}{}

\makeatletter

\begin{document}

\title{Incorporating Individual and Group Privacy Preferences in the Internet of Things}


\author{Khaled Alanezi         \and
        Shivakant Mishra 
}


\institute{Khaled Alanezi \at
              Computer Department \\
              College of Basic Education \\
              PAAET, Kuwait\\
              \email{kaa.alanezi@paaet.edu.kw}           
           \and
           Shivakant Mishra \at
           Computer Science Department \\
           University of Colorado, Boulder, CO, USA.\\
           \email{mishras@cs.colorado.edu}}

\date{Received: date / Accepted: date}

\maketitle

\begin{abstract}

This paper presents a new privacy negotiation mechanism for an IoT environment
that is both efficient and practical to cope with the IoT special need of
seamlessness. This mechanism allows IoT users to express and enforce their
personal privacy preferences in a seamless manner while interacting with
IoT deployments. A key contribution of the paper is that it addresses the privacy concerns of individual
users as well as a group of users where privacy preferences of all individual users are combined into
a group privacy profile to be negotiated with the IoT owner.
In addition, the proposed mechanism satisfies the privacy
requirements of the IoT deployment owner. 
Finally, the proposed privacy mechanism is agnostic to the actual IoT 
architecture and can be used over a user-managed, edge-managed or a
cloud-managed IoT architecture. Prototypes of the proposed mechanism have
been implemented for each of these three architectures, and the results show the capability of the protocol to negotiate privacy while adding insignificant time overhead.
\end{abstract}

\section{Introduction}
\label{sec: introduction}

With increasing availability of wide range of commercial IoT systems nowadays, we are closer than ever into realizing the IoT vision of sensing, interconnecting, actuating and remotely controlling physical objects in our environment. Users can now shop for smart thermostats, door locks, light bulbs, home security systems or patient monitoring systems among other options and control them from their mobile devices. While such systems can bring great comfort in our everyday lives, there is a pressing need to consider user privacy in designing and using them. We see that addressing the privacy concerns of such systems involves answering a two-fold question. First, the simpler question (though answering this question in itself is not simple) is how to allow an individual user who comes across an IoT environment to be in charge of what information can be collected about them? For example, can a user ensure that a camera system in the environment may take her picture but must blur her face before distributing it to any third party, or the IoT environment must anonymize her location sufficiently before sharing it with anyone else. While determining and respecting individual user privacy requirements is complex, a much more sophisticated situation arises when a group of multiple users immerse in an IoT environment. Each user in the group may potentially have a different set of privacy concerns, and the complex question is how to determine the privacy requirements of the group as a whole (group privacy) and provide a service that respects these group privacy requirements. For example, consider a situation where multiple users walk into a room equipped with a video surveillance system. While some users in the group are comfortable with the video capture process, others might require specific policies to be applied to the captured videos, such as face filters or guaranteeing non-disclosure of the video to third parties.  In this journal paper, we extend our previous work {\citep{alanezi2018privacy}} that focused on individual user privacy preferences to incorporate group privacy preferences.

To cater to the privacy preferences of users whether in individual or group context, we propose a privacy negotiation mechanism to allow IoT users to automatically express and enforce their personal privacy preferences while interacting with IoT deployments. The proposed negotiation mechanism is designed to be practical by allowing users' devices (e.g. their smartphones) to negotiate with the IoT deployment in background on users' behalf, thus allowing them to express their privacy requirements without having to go through the burden of explicitly specifying their requirements on an online form or reading privacy notices. Further, this privacy negotiation mechanism is designed to be efficient as it is built on top of existing IoT communication protocols and adds only a negligible overhead to the ongoing communication of IoT applications. A key novelty of this work stems from demonstrating through practical experiments the viability of automated privacy negotiation over IoT architectures.

This proposed negotiation mechanism is holistic in nature in that it not only covers the privacy requirements of the IoT users (i.e. IoT service consumers) but extends this coverage to negotiate and satisfy the privacy requirements of the IoT deployment owner (henceforth IoT owner) as well. The IoT owner is the responsible party for setting up and maintaining the IoT infrastructure that provides services to IoT users. Generally, providing services involves enabling the IoT users to access sensory interfaces such as temperature, audio or camera sensors. We envision that the IoT owners would like to manage this access according to their privacy requirements since every access to a sensor can 
be used to reveal information about them. To cover the privacy requirements of IoT users and owners, we model the problem as a utility-privacy tradeoff function in which sharing more information increases the utility gained from the service but can potentially lead to some undesired privacy exposure. We show that the negotiation protocol can use multiphase negotiation efficiently to embark into an agreement that satisfies the utility and privacy requirements of IoT users and owner simultaneously.

Another important aspect that is covered by the negotiation protocol is the handling of current diverse IoT  infrastructures. IoT infrastructures are envisioned to act as utility networks for IoT users to use IoT services on the go \citep{stankovic2014research}. However, the gateway to access these services can vary from a privately-owned central server or cloud server to an edge server or even direct access using machine-to-machine (M2M) communication. An important feature of our negotiation protocol is that it is architecture agnostic and can be used on cloud based, edge-based or M2M types of infrastructure. 
In summary, our contributions are as follows:

\begin{enumerate}
\item  We have designed a negotiation protocol to automatically model and realize privacy requirements in an IoT ecosystem. The protocol takes a holistic approach by covering the requirements of all involved parties in an IoT interaction. The practicality of the protocol stems from modeling privacy as a tradeoff function with the utility achieved from using/providing IoT services.
\item The proposed negotiation protocol addresses the privacy requirements in both individual and group contexts. In particular, we extend the individual privacy negotiation protocol we proposed in {\citep{alanezi2018privacy}} to address group privacy preferences. In group negotiation, privacy preferences of each individual user are combined into a group privacy profile to be negotiated with the IoT owner. The outcome of the group negotiation could be either to proceed with the data collection, alter content to accommodate for a privacy restriction from group members or to contact one or more of the group members to perform a second round of negotiation.
\item  We have developed and tested a prototype of the proposed negotiation protocol for both individual and group contexts over three different IoT infrastructure standards of user-managed, edge-managed and cloud-managed IoT architectures. Results show the capability of the protocol to negotiate privacy preferences in both individual and group contexts while adding insignificant time overhead. 
\end{enumerate}

\section{Related Work}
\label{sec: related}
There is a consensus among researchers that difficulty in preserving user privacy is a major hurdle for wide adoption of IoT \citep{ziegeldorf2014privacy, naeini2017privacy, stankovic2014research}. Consequently, many researchers proposed solutions to address privacy concerns inherent in IoT systems such as background collection of sensory data without user consent and possibly sharing this data with untrusted third parties such as cloud providers. This literature review focuses on describing the body of work that was proposed to face these challenges. The UPECSI framework \citep{henze2014user} proposed privacy enforcement points at selected network locations where sensitive IoT data must be encrypted before shipping to the cloud. Users are in charge of choosing the privacy enforcement option to be applied at these network points. The privacy coach \citep{broenink2010privacy} is also a framework that puts the control of the data being collected in the hand of the user. The framework is geared towards RFID systems and provides a mobile client where users can review and approve data collection policies proposed by corporations that own the RFID system. Since Blockchain (BC) technology eliminates the need for a central authority to establish trust it was used in \citep{dorri2017blockchain} to allow an IoT owner to define an access control list of acceptable peer-to-peer communications that can take place with their owned devices. The challenge addressed in this work was to build a BC architecture suitable for low resource and limited energy IoT devices. The position paper in \citep{davies2016privacy} presented an architecture based on edge computing to allow the user to control access to their data in an IoT system. The authors argued that the use of cloudlets benefits the IoT system by gaining the trust of the user. Our framework adopts a similar approach to these works in placing the control of IoT data privacy at the hand of the user. However, we address the challenge of enforcing the user privacy policy seamlessly (i.e. without user involvement), which is a key for the usability of any framework to serve privacy in IoT .  

Another challenge addressed by our work is to adopt to privacy preferences of group of users that are subject to a data collection process by an IoT deployment. Requiring privacy for a group is a commonplace scenario in IoT and we focus in our work on building a generic architecture to support this requirement. To the best of our knowledge, we are the first to study this problem. Hence, we cover in this literature other works that considered group privacy in different domains.

Applying group privacy for multi-owned data items (e.g. pictures) shared on social networking sites SNS is a widely studied subject \citep{squicciarini2009collective, such2016resolving, thomas2010unfriendly, such2016privacy, hu2012multiparty, fogues2017sosharp, lampinen2011we}. The authors in these works looked into automatically identifying co-ownership and practically applying group privacy on such items. 

According to the author in \citep{mittelstadt2017individual} another domain where privacy must also be considered in a a group context is big data analytics. The author of this work proposes that when users are categorized among groups automatically by big data algorithms, there is an ethical responsibility that information and actions linked to the group must be considered with group privacy in mind. 

\section{Scenarios}
\label{sec:scenarios}

We identify three IoT scenarios that are common
in literature namely user-managed IoT, edge-managed IoT
and cloud-managed IoT. In a user-managed IoT architecture, the IoT owner's
mobile device, e.g. a smartphone or a tablet acts as a gateway
to the IoT infrastructure. All negotiation and access
must happen through this mobile device. This is common
in an IoT infrastructure installed at home or a private office.
Consider a smart home environment where a home is equipped with a range of
sensors such as a temperature sensor, a carbon monoxide sensor, etc. These
smart home sensors are designed to operate in a star topology network, where
each sensor is connected to a single gateway, e.g. home owner's smartphone,
and sends its data to that gateway at regular intervals. 
In this case, if another device, e.g. a visiting friend's smartphone needs
to access this sensor data, possibly for a limited time, the sensor data 
must be retrieved from the home owner's smartphone. In this scenario, the
home owner's smartphone must be involved in privacy negotiation with the
 smartphone of the visitor to permit the temporary access.

For edge-managed and cloud-managed IoT, a server plays
the role of the gateway and carries all negotiation and access
requests. For the former, the server is within the same network
domain for the IoT infrastructure\citep{davies2016privacy}, whereas for
the latter, the server is situated in a public cloud. For example, in a 
smart city, video feed from a security camera may be stored at an edge server 
or a cloud server. If an agency such as a law enforcement agency needs to
access this video feed, the edge or cloud server must be involved in privacy negotiation with the law enforcement agency.
 
These three scenarios categorize the negotiation process from the perspective of the owner. On the other side, when it comes to the IoT users, there are two possible scenarios that could take place. Either an individual user is negotiating with the owner to ensure that his\slash her privacy preferences are met or a group of users are present with different privacy preferences pertaining to each individual in the group. The latter case is more complex since individual preferences of different users may conflict with one another in some situations. We present a negotiation algorithm covering each of these scenarios in Section \ref{sec:design}.

\section{Design}
\label{sec:design}
\subsection{Privacy Requirements}
\label{subsec:design-req}
The fact that the negotiation protocol operates in IoT environments imposes stringent practical requirements on our design. We describe below four of these requirements that must be satisfied: 
\begin{enumerate}
\item \textbf{No User Involvement:} All negotiation communication must take place in background, without any user intervention. When users navigate through public or private spaces, they will typically encounter IoT deployments (i.e. IoT owners) and engage in data exchange with them to avail various services provided.  To avoid any inconvenience introduced by involving IoT users in negotiation to satisfy their privacy requirements, the IoT user's device that has the user privacy requirements should act seamlessly to negotiate these privacy requirements on behalf of the user with the IoT owner. Also, the IoT owner's privacy settings will be communicated to the IoT user to ensure their adherence to these requirements.
\item \textbf{Minimal Overhead:} The negotiation protocol must impose minimal time and energy overhead to the IoT task. IoT services are typically provided promptly to the users. Therefore, these services are sensitive to any time delays, which should be considered in the protocol design. Also, since IoT devices are battery powered, the protocol must be energy efficient to avoid draining the energy sources of involved devices.
\item \textbf{Choice Flexibility:} The protocol should avoid the current privacy notice and choice model of either accepting the service as a whole or abandoning it, since this model is too rigid and will not work with IoT situations requiring more flexibility. Diverse options should be offered to IoT users to enable them to continue to use the service while not sacrificing their privacy requirements. 
\item \textbf{Guaranteed Privacy:} Regardless of whether the negotiation is happening in an individual or group context, the protocol shall never proceed with a data collection action or providing a service that violates an individual privacy preference. In the case of individual context, if the IoT user and the IoT owner did not reach an agreement, any suggested alternative data collection policy along with a differentiated service by the IoT owner must be listed as an acceptable option by the individual (i.e. a plan B option). For the group context, individual group preferences will be aggregated using the least misery approach to ensure satisfying the user with most stringent privacy preference. After that, the group negotiation with the IoT owner will use the resultant privacy preference from the aggregation process. In cases where no agreement can be reached, the corresponding user(s) will be notified with a suggested policy from the IoT owner to obtain their approval in a similar fashion to the individual negotiation.
\end{enumerate}
\subsection{Privacy Model}
\label{subsec:privacy-model}
The negotiation protocol presented here is in alignment with the vision of achieving openness in IoT environments. Openness envisions IoT networks to act as a utility infrastructure, similar to electricity and water, that is accessed by IoT users on the go \citep{stankovic2014research}. Open environments as such require that the parties involved in information exchange specify their privacy requirements to be negotiated on their behalf. This is analogous to the P3P protocol \citep{cranor2002web} in which a browser negotiates the privacy requirements of users on their behalf with visited websites to control the personal information that the website can collect about the user. In our protocol, the IoT user and IoT owner will store their privacy requirements in a policy file  stored locally and written using XML language. An example scenario for privacy policies for an IoT owner and an IoT user is shown in Listing 1 and Listing 2 respectively. 

\begin{lstlisting}[frame=single,basicstyle=\small,language=XML,caption=IoT User Privacy Policy\label{code:user-policy}]
<privacy-policy>
  <data-in type="image" priority="1">
    <retention>3-month</ retention >
    <shared>no</shared>
    <inferred>yes</inferred>
  </data-in>
  <data-out>
  <data-out type="video" priority="1">
    <retention>1-year</ retention >
    <shared>no</shared>
    <inferred>no</inferred>
  </data-out>
</privacy-policy>
\end{lstlisting}

\begin{lstlisting}[frame=single,basicstyle=\small,language=XML,caption=IoT Owner Privacy Policy\label{code:owner-policy}]
<privacy-policy>
  <data-in type="video" priority="1">
    <retention>1-year</ retention >
    <shared>no</shared>
    <inferred>yes</inferred>
  </data-in>
  <data-out type="face-detection" 
  priority="1">
    <retention>1-year</ retention >
    <shared>no</shared>
    <inferred>no</inferred>
  </data-out>
  <data-out type="image" priority="1">
    <retention>1-year</ retention >
    <shared>no</shared>
    <inferred>yes</inferred>
  </data-out>
</privacy-policy>
\end{lstlisting}

As seen in Listing 1, the privacy policy for the IoT user specifies \textbf{<data-in>} tags indicating the type of data that the user would like to acquire from the IoT owner along with child elements specifying the usage scenario for this data. These \textbf{<data-in>} tags from the IoT user privacy policy will be matched against the \textbf{<data-out>} tags in the IoT owner's policy since the latter specifies the data collection practices accepted by the IoT owner. Conversely, the \textbf{<data-in>} tags specified in the IoT owner's policy in Listing 2 will be matched against the \textbf{<data-out>} tags in the IoT user  privacy policy to ensure that the level of data collection performed by the IoT owner is acceptable by the IoT user. An interesting research question is how to learn the privacy policy for each user which could be different for different locations? Other research \citep{naeini2017privacy} has shown that the privacy preferences of a user can be predicted by observing a few data collection scenarios. Using prediction is beneficial to avoid the cumbersome and error prone task of filling privacy settings screens.
One of our key design requirements is achieving flexibility by allowing the protocol to negotiate multi-levels of service with different scales of data collection scenarios. This is required due to the high cost of an unsuccessful negotiation as it might require the user to leave the place to avoid the data collection process altogether. To address this challenge, the negotiation protocol models the relationship between data collection and the IoT service as a utility-privacy tradeoff function stated as Formula 1. This allows the IoT owner to offer multiple choices of service levels, measured by the utility, to the IoT user based on the amount of data they are willing to share. We choose to employ four dimensions form factors influencing privacy preferences in IoT environments from \citep{naeini2017privacy} in this utility-privacy tradeoff function. 
These four factors are saved in the privacy policy XML representation as child elements inside each \textbf{<data-in>} and \textbf{<data-out>} tags. The description for each of these four elements is as follows:
\begin{enumerate}
\item \textbf{Data Type (\it t).} The type of sensor being accessed can have varying degree of exposure to the privacy of the owner of that sensor. Sensors such as the camera or the microphone are inherently sensitive. Hence, allowing access to those sensors must be handled with care. There are techniques in literature to minimize the degree of privacy exposure when accessing those sensors such as blurring faces from the live video feed of a surveillance camera \citep{das2017assisting} or carefully choosing audio features to avoid construction of speech from captured audio data \citep{wyatt2007conversation}. If used, these techniques must be added to the XML file to be part of the negotiation process.
\item \textbf{Retention. (\it r)} Retention policy specifies time durations for keeping logs of exchanged data. In real-time applications, where no data storage is required, this factor can be used by the IoT owner or the IoT user to enforce purging their data by leaving this element empty.
\item \textbf{Shared (\it s).} Any third party recipient must be specified in case the IoT owner or the IoT user is sharing any gathered data for the IoT task.
\item \textbf{Inferred (\it i).}  The recipient of the data must specify if inference techniques will be used to gain further information from the data. For example, accelerometer data can be used to monitor exercising habits of a user for health applications but can also be used in dead reckoning techniques \citep{pratama2012smartphone} to determine indoor user location.
\end{enumerate}
Other form factors that can also be considered include the location, purpose and the benefit of the data collection \citep{naeini2017privacy}. After learning the aforementioned privacy influencing factors in the negotiation exchange described in next two subsections, both the IoT owner and IoT user will use them as part of a privacy-utility calculation to ensure that the achieved utility from the IoT service outweighs the degree of privacy exposure. We adopted the privacy-utility function in \citep{preibusch2006implementing}, which serves electronic commerce (e-commerce) sites, with modifications to fit to IoT applications scenario. This function is used by both the IoT user and the IoT owner of the data to evaluate the privacy-utility of the the data exchanged between them. The utility-privacy function is as follows:
\begin{equation}
U = -\gamma.P_e(t,r,s,i) + B(t,r,s,i)\label{eqn:individual}
\end{equation} 


\textit{U} denotes the total utility that will be achieved from pursuing the information exchange to run the IoT application.  

\textit{B} is the benefit from the data exchange as seen from the perspective of the data owner. For the IoT owner, this could be a monetary incentive or the social benefit from allowing IoT applications to run on their premises. As for the IoT user, the benefit would be the service provided by the IoT application. Notice that \textit{B} is a function of privacy exposure form factors as we expect the benefit to be proportional to the selected policy for each data type. Therefore, we can model \textit{B} as the sum of the benefits achieved by choosing the policy  configuration for each policy item (i.e. \textit{t,r,s,i})  and finding the summation of benefits for all data types involved in the data exchange scenario as follows:

{\small
\begin{eqnarray}
\sum_{i=1}^n [B(t_i)+B(r_i)+B(s_i)+B(i_i)]
\end{eqnarray}
}

\textit{$P_e$} is the degree of privacy exposure for the selected privacy policy. Different privacy policies will lead to different levels of privacy exposure based on the exposure form factors chosen in the policy. For example, higher retention periods specified by \textit{r} for highly sensitive data specified by \textit{t} will lead to higher values of \textit{$P_e$}.

In contrast with the benefit \textit{B}, we choose to model the privacy exposure \textit{$P_e$} as the product of privacy exposure terms resulting from the selection of policy items and finding the summation of \textit{$P_e$} across the data types as follows:

{\small
\begin{eqnarray}
\sum_{i=1}^n [P_e(t_i)*P_e(r_i)*P_e(s_i)*P_e(i_i)]
\end{eqnarray}
}

The design choice of using products for modeling \textit{$P_e$} stems from the fact that different configuration parameters can affect each other with direct proportionality. For example, choosing a three month retention period for a video will be of higher privacy concern than using the same retention period with a less sensitive sensor such as the accelerometer. 

We also note that the degree of privacy exposure for each data type can vary based on the user location context (private vs. public place) and the social context (individual vs. group). Since this privacy perception varies from one user to another, user questionnaires are proposed \citep{naeini2017privacy} to learn the individualized user preferences that can then be used as policy parameters.

\textit{$\gamma$ } is an overall privacy sensitivity perception factor. This factor can vary depending on the location or context of the user. For example, users might be comfortable for taking a picture for them in public places as opposed to being in private places. \textit{$\gamma$ } is multiplied by \textit{$P_e$} to either escalate or deescalate the total privacy leakage for the specific data sharing situation based on the context and/or location. 

Note that the product term to the right of the equality operator is negative so as to reconcile it with the benefit term \textit{B}. Thus, the utility \textit{U} will be positive when the value of the  benefit \textit{B} term outweighs the value of the negative privacy exposure term and vice versa.

In summary, the presented privacy model uses XML to store the privacy policies and use them without user intervention. Also, it supports the requirement of choice flexibility by modeling the utility of the service as a function of privacy exposure form factors. Section \ref{sec:evaluation} demonstrates that the model and the overall protocol satisfies the minimum overhead requirement when implemented over well known IoT architectures.

\begin{figure*}
\centering
\subfloat[1-Phase Negotiation]{\includegraphics[width=2in, height=2.6in]{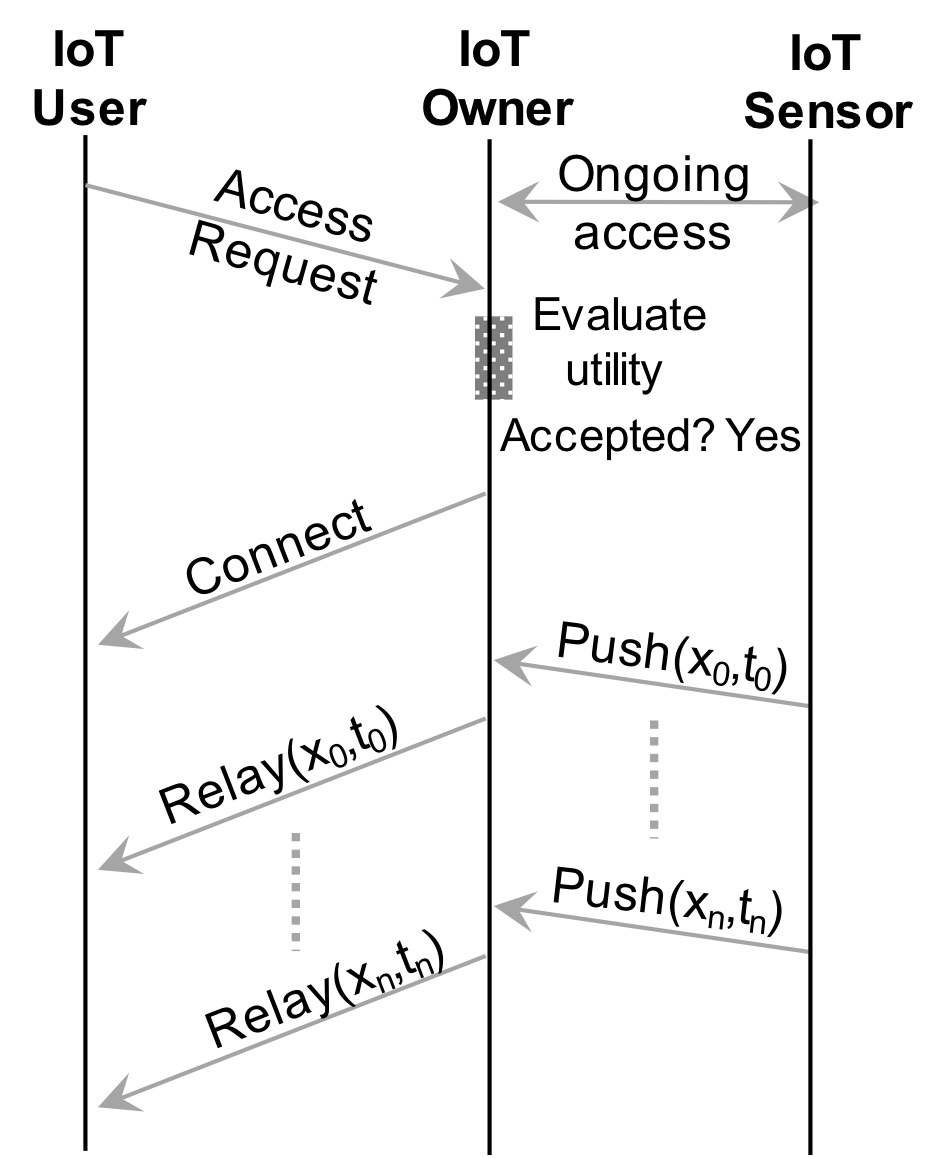}}
\hspace{1in}
\subfloat[2-Phases Negotiation]{\includegraphics[width=2in, height=2.6in]{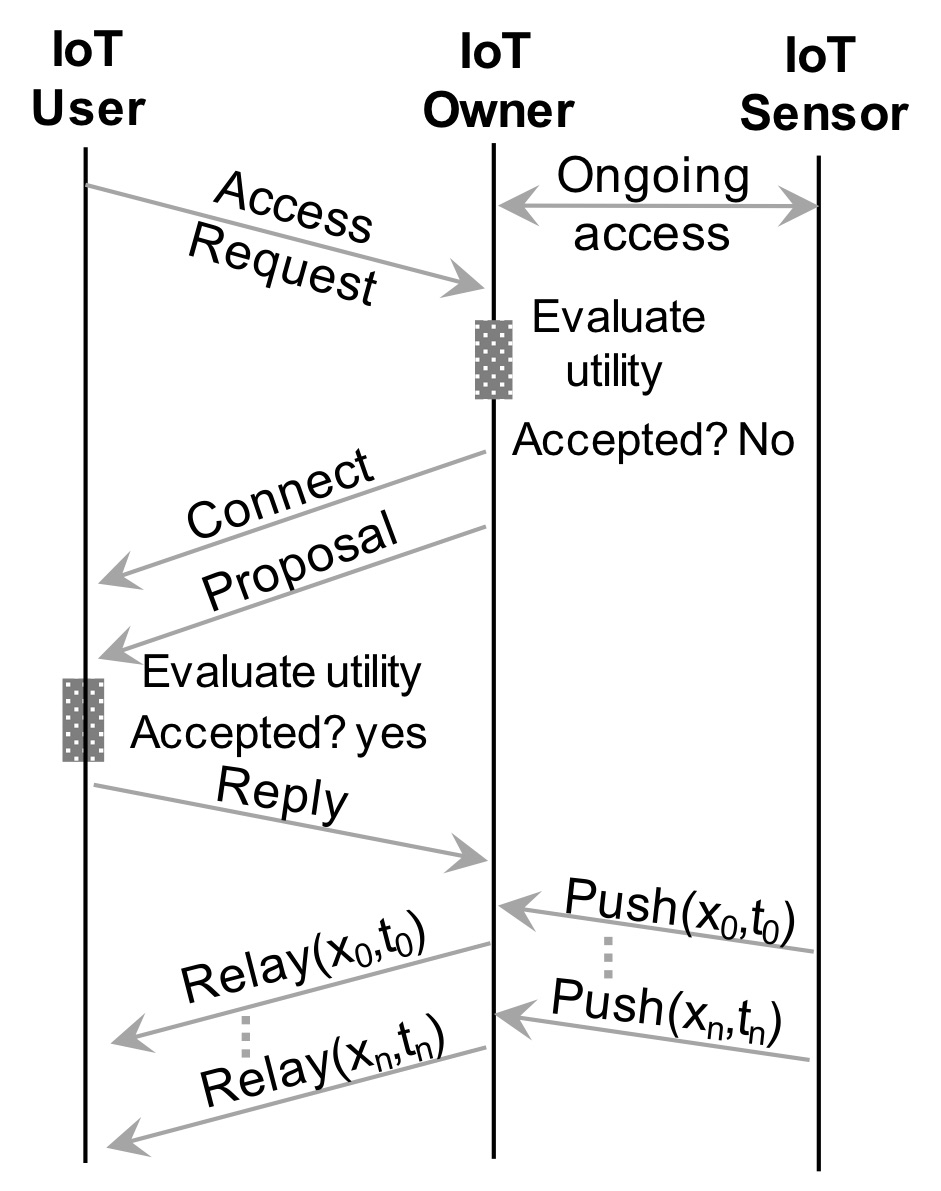}} 
\caption{Privacy Negotiation Flow}
\label{fig:experiment_setup}
\end{figure*}
\subsection{Individual Privacy Negotiation}
\label{subsec:privacy-negotiation}
This section describes the flow of the negotiation activities between the IoT user and the IoT owner in the individual context (i.e. when there is a single user in the IoT environment). We have implemented and evaluated the privacy negotiation protocol on all three architecture scenarios described in Section~\ref{sec:scenarios}.  
The design of the negotiation protocol is made flexible by avoiding the go/no-go scheme of current privacy/notice choice mechanisms and replacing it with multi-phase negotiation. This section describes 1-phase negotiation and 2-phases negotiation. We start by describing the 1-phase negotiation scenario shown in Figure \ref{fig:experiment_setup} (a). The scenario starts with the IoT user sending an access request for specific type of data (i.e. sensor) to the IoT owner. The IoT user will embed a summary of their usage requirements for this data, which is taken from the \textbf{<data-in>} element for this type of data in the user policy. Upon receiving the access request, the IoT owner checks the utility of the request by substituting it in the utility function. Assuming that the utility of the request is equal or higher than the utility achieved when substituting the \textbf{<data-out>} element for the same data from the IoT owner privacy policy, the owner accepts the request. After that, the IoT owner connects to the IoT user and starts acting as a relay by forwarding sensor information received from the IoT infrastructure to the IoT user.
The 2-phases negotiation is shown in Figure \ref{fig:experiment_setup} (b). This scenario starts in a similar way by the IoT user sending an access request to the IoT owner. However, in this scenario the utility of the request is deemed unacceptable by the IoT owner. Accordingly, the IoT owner will retrieve the \textbf{<data-out>} element for the requested data item from their own privacy policy, which represents their acceptable usage scenario for the data item and send it as a proposal to the IoT user. This only happens after connecting to the IoT user. The IoT user's device then checks the utility of this proposal against a second priority policy if one is defined in their privacy policy file. Note that each \textbf{<data-in>} and \textbf{<data-out>} tag in the privacy policy contains a priority attribute to allow the IoT user and IoT owner define alternative policies to be used during negotiations. Only defining a priority 1 policy for the data item means that the policy for this data item is non-negotiable. Assuming the IoT user has accepted the alternative proposal, the IoT owner will start forwarding required data as soon as it is received from the data source.

\subsection{Group Privacy Negotiation}
\label{subsec:group-privacy-negotiation}
The previous section focused on performing negotiations between an individual user and the owner of the IoT environment. However, an important question is how to perform negotiations when there are multiple users present in the scene? Answering this question is vital since IoT deployments cover urban spaces such as malls, hospitals, ...etc., where the presence of multiple users interacting with the IoT deployment is the norm.

Let's consider a group of $n$ users $U=(u_1,\cdots, u_n)$ who happen to be at the same location where an IoT deployment exists. For the IoT environment to provide a particular service, it requires performing data collection of one or more sensors represented by the vector $S=(S_1,\cdots, S_k)$. The privacy setting of each user is specified per sensor as follows:\\  $S_{ps}=(<S_1, r, s, i>,<S_2, r, s, i>,\cdots,<S_k, r, s, i>)$ \\
 This presentation is similar to the notation used in the individual privacy negotiation as described in Section \ref{subsec:privacy-negotiation}. We expect that specifying user privacy settings at the sensor level would be easy for users as it is analogous to the way users specify their privacy settings when granting permissions to mobile applications at the time of installation in their smart devices.

Indeed, it may be difficult for users to feed the system with their privacy settings for every sensor before they start using the system.  However, prediction based on limited user input can be used to learn these settings  effectively \citep{naeini2017privacy} thereby minimizing required user efforts. Also, we believe that spatiotemporal context as well as social group context (with family, friends, colleagues, ..etc) can be a huge influencer for user choices and should be considered while predicting a user privacy settings for a particular location.

Given that all available users will post their privacy settings to IoT service owner in order to consider them in the data collection policy, the IoT service owner will end up with a matrix of group preferences with each row representing the requirements for an individual user as shown in Table \ref{fig:group-table}. Note here that the individual rows contain the form factors (i.e. r, sh, i) for each sensor required by the IoT service as specified by each user. 

\begin{table}[]
\begin{tabular}{|l|l|l|l|l|l|l|l|}
\hline
\rowcolor[HTML]{EFEFEF} 
                            & \it{S1}r & \it{S1}s & \it{S1}i & \it{S2}r & \it{S2}s & \it{S2}i & ... \\ \hline
\cellcolor[HTML]{EFEFEF}U\textsubscript{1}  & 3   & yes & yes & 3   & no  & yes & ... \\ \hline
\cellcolor[HTML]{EFEFEF}U\textsubscript{2}  & 4   & no  & yes & 3   & no  & no  & ... \\ \hline
\cellcolor[HTML]{EFEFEF}... & ... & ... & ... & ... & ... & ... & ... \\ \hline
\cellcolor[HTML]{EFEFEF}U\textsubscript{n}  & 2   & yes & no  & 3   & no  & yes & ..  \\ \hline
\cellcolor[HTML]{EFEFEF}Boundary & $lm_r$ & $lm_s$ & $lm_i$ & $lm_r$ & $lm_s$ & $lm_i$ & ... \\ \hline
\end{tabular}
\caption{Group Privacy Policy Matrix and Boundary Preference}
\label{fig:group-table}
\end{table}

The last row in the table contains the boundary policy resulting from merging the privacy preferences of the group of users. We choose to merge the privacy preferences using the least misery approach to ensure that no information can be disclosed without the consent of all users. Using this approach, the resulting boundary privacy preferences vector will contain in every item a privacy setting that satisfies the group member with the most conservative privacy requirement. For example, if $U_1$ specifies that data from $S_1$ can be stored for maximum 3 months while $U_2$ requires that the maximum is 1 month only, the aggregate policy will list that the data item $S_1$ will be stored for maximum 1 month.

The boundary privacy policy can be used as the policy of choice as it does not violate the privacy requirements of any user. However, such a choice for the design would be naive considering that the algorithm should guide the IoT users in balancing the utility of the service consumed with the amount of privacy exposure the users will be subjected to. For example, one possibility that the algorithm should not ignore is the availability of  a policy that is more conservative than the boundary policy that shall provide a service utility that still satisfies the users. Hence, the aggregate privacy policy will be chosen by enumerating all possible combinations of form factors that are within the boundary privacy preferences vector and then again plugging every possible combination into an optimization problem where the goal is to minimize the data collected about the users (privacy exposure) while maximizing the benefits for the users (service utility) as shown in the below equations.

{\small
\begin{eqnarray}
\min_{x \succeq 0}~(\mbox{w.r.t}~R^n_+)&&  (\sum_{i=1}^n Pe_i P_i x_i)\label{eqn:group-pe}
\end{eqnarray}
}

{\small
\begin{eqnarray}
\max_{x \succeq 0}~(\mbox{w.r.t}~R^n_+)&&  (\sum_{i=1}^n B_i x_i)\label{eqn:group-b}
\end{eqnarray}
}

Equation \ref{eqn:group-pe} in the optimization problem represents the sum of the privacy exposure terms $Pe_i$ caused by collecting an amount of data $x_i$ under a privacy policy $P_i$. The goal of the optimization is to maximize this term. Meanwhile, the optimization will maximize equation \ref{eqn:group-b} which represents the sum of the benefits $B_i$ caused by collecting the same amount of data $x_i$ under the same privacy policy $P_i$. These equations add the the privacy policy term $P_i$ to equation \ref{eqn:individual} to cater for evaluating and choosing among various privacy policies. 

Notice that in the first round of the negotiation, the optimization will only consider the set of policies that do not violate (i.e. work within the limits of) the boundary privacy policy items derived using the least misery policy. However, it is important to consider situations in which one tight restriction from a user might render the service unusable for the entire group. For example, a user refusing to ship some specific data item that must be processed on the server side or by a third party. Given that the user is available in the IoT domain, the algorithm can contact them to see if they are willing to compromise in return for a better service. Finally, if there is no agreement reached in the second negotiation phase, the IoT owner can send a notification to the user device telling the user that their privacy requirements can not be met and providing them with detailed information about the data collection that will take place for them to decide to continue or to simply leave the location.

\section{Implementation}
\label{sec: implementaion}
In a user-managed architecture, the dominant communication protocol is BLE to conserve energy, while for the edge-managed and cloud-managed architectures, a server acts as a gateway and communication is done using WiFi. Hardware used to simulate these infrastructures is shown in Table \ref{table:hardware}. 
\begin{table*}
\footnotesize
\caption{List of hardware components used in the experiments along with the corresponding function and the scenario for each component.}
\begin{center}
\rowcolors{2}{gray!25}{white}
\begin{tabular}{ |c|c|c| } 
\rowcolor{gray!50}
  \hline
 Title & Function & Scenario Used \\
  \hline 
Arduino Uno R3 & Microcontroller for the IoT device & User managed and edge-managed \\ 
LM35 Temperature Sensor & Sensory interface to be accessed & User managed and edge-managed \\ 
Arduino Ethernet Shield & Adds ethernet connectivity to the Arduino Uno  & Edge-managed and cloud-managed \\ 
Arduino BLE Shield & Adds BLE connectivity to the Arduino Uno & User-managed \\ 
D-Link Wireless Router & Wireless LAN router & Edge-managed and cloud-managed \\ 
Motorola Moto E smartphone & IoT user requesting access to a sensor & All \\
Motorola Moto E smartphone & IoT owner & User-managed \\
MacBook Air & Computing node at the edge of the network & Edge-managed \\
Amazon EC2 Micro Instance & Computing node in the Cloud & Cloud-managed \\
Arduino Uno ESP2866 & Microcontroller for the IoT camera & Edge-managed/Group Privacy \\
Arducam 2MP Camera & IoT camera for generic use  & Edge-managed/Group Privacy \\
 \hline
\end{tabular}
\end{center}
\label{table:hardware}
\end{table*} 

\subsection{User-Managed IoT Experiment}
\label{sec:person-managed}
\begin{figure*}
\centering
\subfloat[User-managed experiment.]{\includegraphics[width=2.1in,height=2.3in]{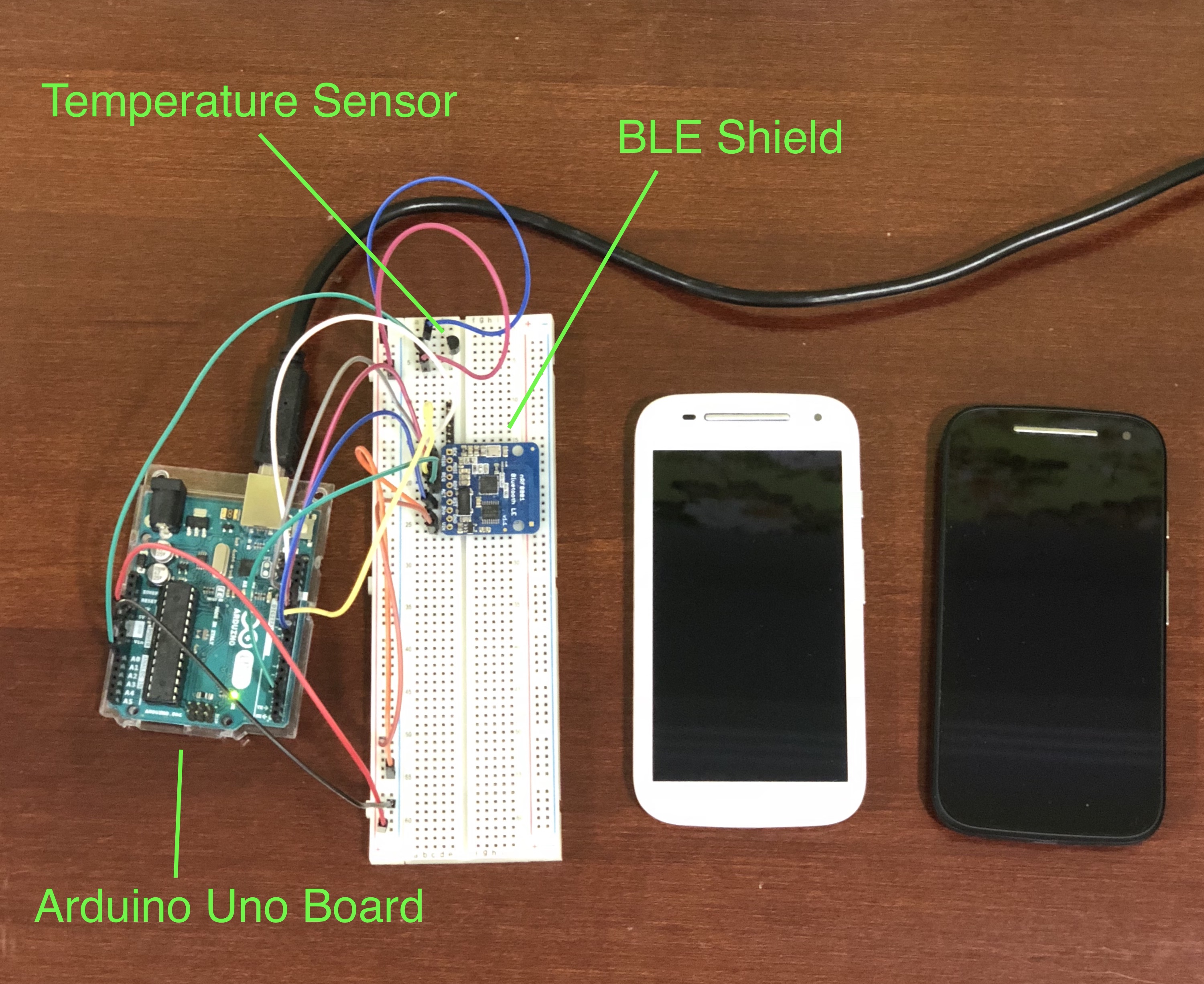}\label{fig:user-managed-exp}}
\hspace{0.1in}
\subfloat[Cloud\textbackslash Edge-managed experiment.]{\includegraphics[width=2.1in,height=2.3in]{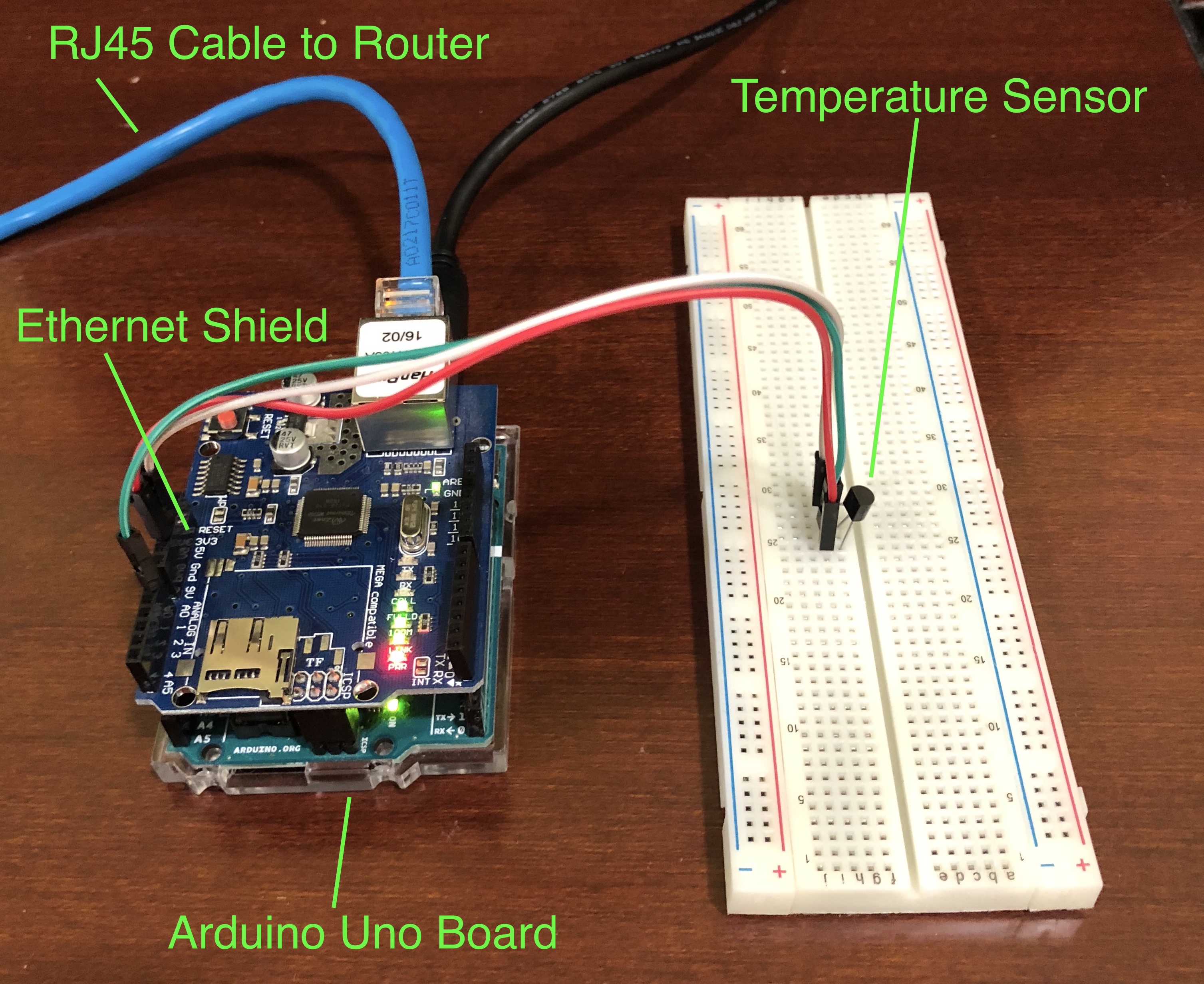}\label{fig:edge-managed-exp}} 
\hspace{0.1in}
\subfloat[Camera group privacy experiment.]{\includegraphics[width=2.1in,height=2.3in]{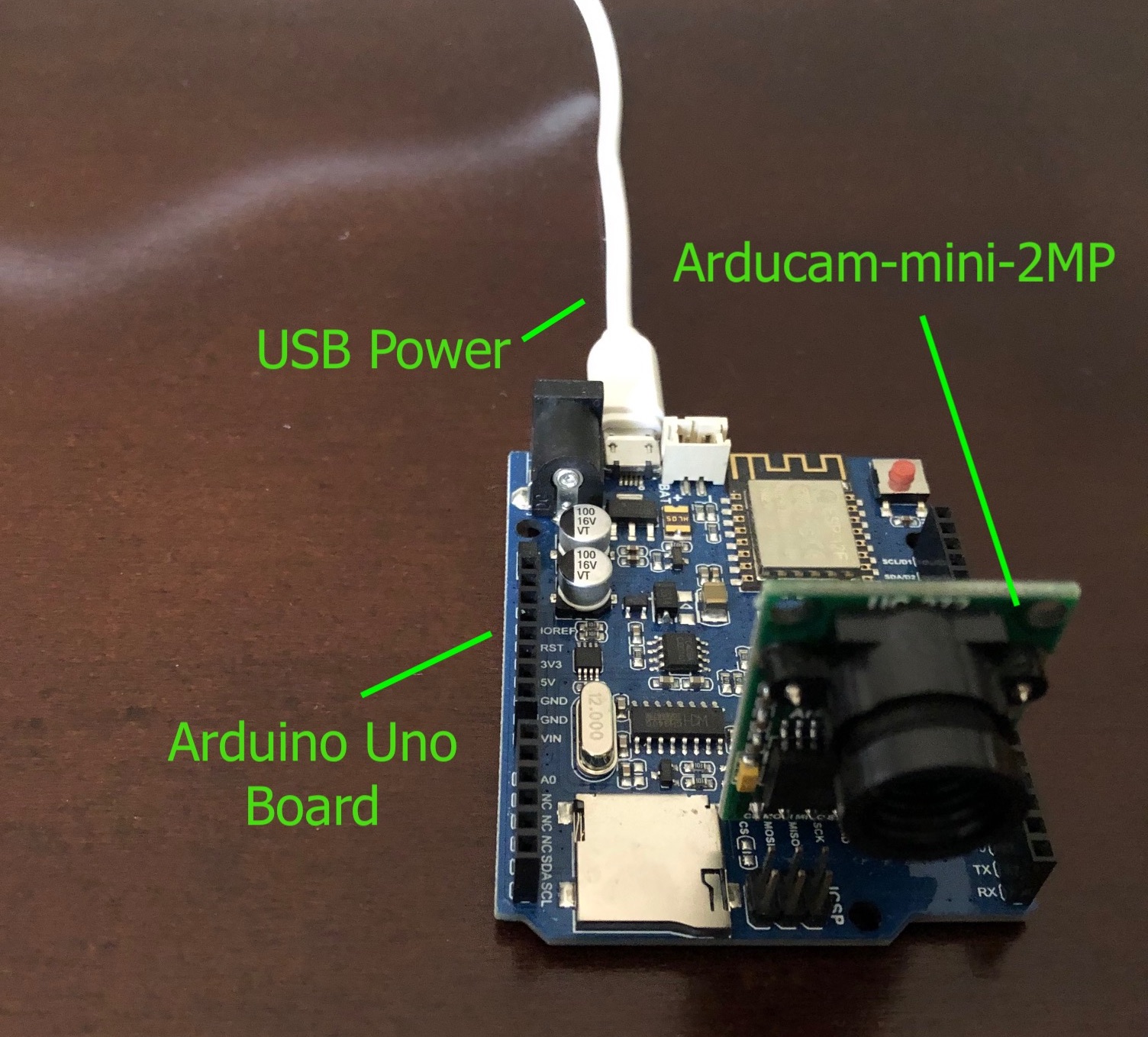}\label{fig:camera-exp}} 
\caption{Hardware Setup.}
\label{fig:experiment_setup_impl}
\end{figure*}

The picture in Figure \ref{fig:user-managed-exp} demonstrates the setup for the user-managed IoT experiment. We used the breadboard to connect a temperature sensor to the Arduino Uno. We then enabled BLE connectivity functionality on the Arduino board by connecting it to an Adafruit Bluefruit BLE shield through the breadboard. Using the Arduino IDE, a program was written to enable the Arduino board to broadcast using BLE beacons its availability in order for BLE clients to connect to it and read the temperature sensor values. The two Android-based Moto E smartphones were then used to simulate the IoT owner and IoT user mobile devices. We wrote Android code to enable the IoT owner to connect to the Arduino board and start receiving temperature values as push notifications periodically. Meanwhile, the IoT owner registers a BLE search to listen to any guest devices that might need a sensor reading from the IoT environment they own. Note here that the BLE search is registered under a specific UUID designated for the sensor sharing services. Searching for specific UUID using BLE can happen in the background while the device is in sleep mode thereby drastically reducing energy consumption. This means that the cost of detecting collaborators for the IoT owner is trivial.

On arriving in an IoT environment, a user who is looking for a specific type of sensor to perform a particular service sends a BLE advertisement broadcasting its intention to access a shared sensor. This BLE broadcast contains the UUID for the sharing service and a vector containing the request information as described in Section \ref{sec:design}. The request information here is a summary from the \textbf{<data-in>} tags describing the data that the device requires along with the usage scenario. This broadcast is captured by the IoT owner as it contains the UUID of the sharing service. In case the IoT owner accepts the request, it connects to the requesting device and starts relaying temperature readings to it as soon as they arrive from the Arduino board. Otherwise, as described in Section \ref{sec:design}, the negotiation flow requires a second round of negotiation. In this case, the IoT owner's device connects to the IoT user's device and sends a proposal containing its acceptable privacy policy for the required sensor. The requester can now either accept or reject this proposal. The requester will use the ongoing BLE connection to reply. If the reply is accept, the owner will start relaying temperature values to the IoT user. Notice that in this situation the IoT owner is a hub for a BLE star topology network with the IoT user and the Arduino board acting as the hosts.  We demonstrate the time efficiency for the communication in the evaluation in Section \ref{sec:evaluation}. If the IoT user rejects the proposal, the owner simply tears down the BLE connection and continues to operate normally.

\subsection{Edge-Managed and Cloud-Managed IoT Experiments}
\label{sec:edge-managed}
We also show in Figure \ref{fig:edge-managed-exp} the setup for the edge-managed and cloud-managed IoT experiments.The Ethernet shield is stacked on top of the Arduino board to provide it with Ethernet capability. After that, an Ethernet cable is used to connect the Arduino board to the wireless LAN router. This allows the Arduino to get a local IP and is now able to communicate with other devices within the same wireless LAN. We used the Arduino IDE to write code to let the Arduino Uno act as web server providing an HTML page to read the temperature sensor values.
This setup is then used to perform 1-phase negotiation and 2-phases negotiation for the edge-managed and cloud-managed IoT scenarios. For the former, we used a MacBook Air laptop to act as an edge server by running Java code listening to network communication at specific port within the same LAN. For the latter, the same Java code was deployed to an Amazon EC2 instance that is used as a cloud server. Port forwarding was used on the wireless LAN router to enable communication between the cloud server and the Arduino board over the Internet. For both scenario, the server (i.e. IoT owner) receives a request from a mobile device resembling the IoT user for sensory data. This request contains a summary from the IoT user policy. The server replies with the sensor information or with a proposal in case further negotiation is required. The client either accepts or rejects this proposal. If the proposal is accepted, the server performs an HTTP GET to get the temperature value and sends the result. We report time-efficiency measurements for the milestone stages for this protocol in Section \ref{sec:evaluation}.

\subsection{Edge-Managed Group Privacy Adaptation Architecture}
\label{sec:edge-group-privacy}

We choose the edge-managed architecture as a basis for evaluating a group-based privacy negotiation scenario. Note that in the other two kinds of architectures of user managed and cloud-managed,  a group-based scenario is also possible. However, for the former the negotiation happens with the IoT owner mobile device whereas for the latter the negotiation happens with the managing cloud server. We choose to implement the edge-based scenario as we believe that edge servers will be a prevalent choice for use as gateways for managing IoT networks. Also note that our implementation focused on developing a generic software architecture that is agnostic to the application or type of sensory data being negotiated and evaluating the impact of the group privacy filter choice on the performance. 

\begin{figure}
  \includegraphics[width=\linewidth]{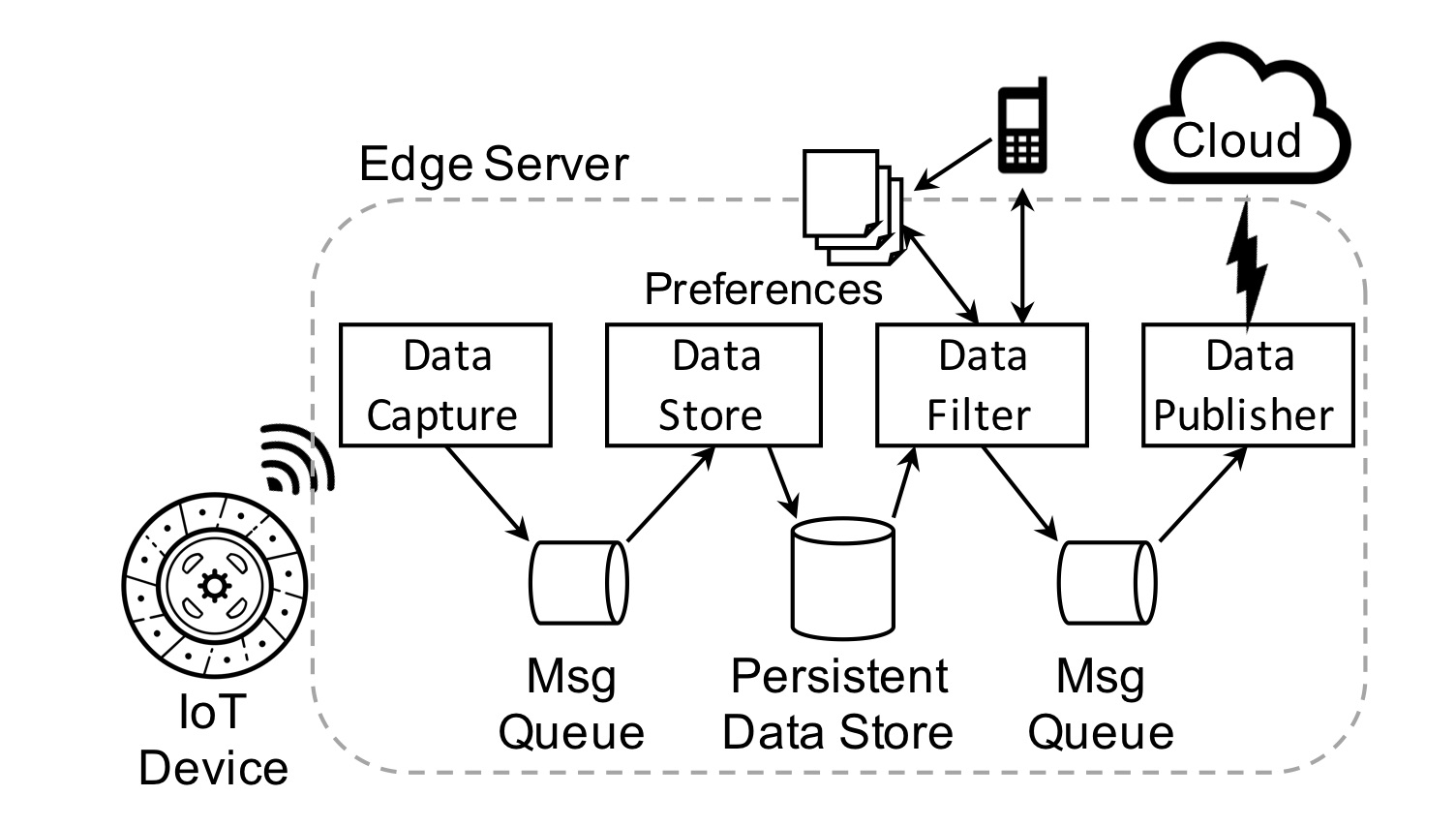}
  \caption{Microservices Architecture}
  \label{fig:group-privacy}
\end{figure}

The situation involving group of users requires activating different kinds of filters depending on the privacy preferences of the group members who happen to be at the scene. We describe in this section a flexible plug-and-play architecture to support this requirement. We choose to implement this architecture using the widely advocated microservices architecture \citep{dragoni2017microservices}. We also adapted a pipeline architecture similar to Eclipse Kura \citep{kura}. By using microservices, it is possible to implement different privacy filters independently (each as a microservice) and bring up/down the filters as required during runtime. Our solution architecture resembling a pipeline structure implemented on the edge server is shown in Figure \ref{fig:group-privacy}. 

As seen in the figure, the overall service is decomposed into four microservices interacting via message queues. Using messages and message queues is necessary in order to achieve independence \citep{dragoni2017microservices}, hence the benefit of the portability that the microservices architecture provides. First, the \textbf{Data Capture Service} starts the pipeline architecture by reading data from the IoT device. The microservices architecture hides the implementation details for the type of communication used between the edge and the IoT devices which can be based on Bluetooth LE or WiFi depending on the IoT device preferred communication protocol. Note that each microservice is implemented using docker a container \citep{docker}. Upon capturing the sensory data, it will be forwarded via the message queue, which is implemented using Java Messaging Service (JMS) \citep{jms}, to the \textbf{Data Store Service}.  The latter is responsible for saving any captured data into a persistent data store for future access. We used docker volumes as a means to implement the persistent data store for the system. The \textbf{Data Filter Component}  is responsible for implementing a privacy filter after consulting the privacy preferences of available users. The framework provides an API for users to submit their preferences immediately upon joining the IoT environment. Received individual preferences are used by the group privacy negotiation protocol described in Section \ref{subsec:group-privacy-negotiation} to achieve an agreed privacy preference vector for the group to apply on the collected data items. Analogous to the individual context, in case one or more members of the group have a conflicting privacy preferences, the privacy filter component can contact the user for an alternative privacy preference (i.e. plan B). We describe a specific data filtering implementation based on camera sensor in Section \ref{subsec:camera-sensor-privacy}. Finally, only filtered data is forwarded to the data publisher, which publishes the data to the cloud to be accessed by other users. The \textbf{Data Publisher} microservice component (or container) is implemented using the desired data communication plugin for the preferred cloud provider (Amazon AWS or Microsoft Azure).

\subsection{Group Privacy Involving Camera Sensor Scenario}
\label{subsec:camera-sensor-privacy}

To evaluate the group privacy architecture, we implemented an IoT scenario involving a group of IoT users negotiating the data collection policy of an IoT camera. The hardware setup for this experiment is shown in Figure \ref{fig:camera-exp}. 

The experiment consists of an Arduino Uno ESP2866 board that is capable of connecting to Wi-Fi interfaced with an Arducam 2MP camera. 
The microcontroller with the camera and the edge server communicate over Wi-Fi to resemble an edge-managed environment. We focused in our experiments on evaluating the effect of applying the privacy filter on the time required to be ready for reporting the image to the cloud. That is, the time from capturing the data until reaching the \textbf{Data Publisher} components. The results of the following three situations are reported in Section \ref{sec:evaluation}:

\begin{enumerate}
\item \textbf{No privacy:} This baseline situation assumes no required privacy filtering, which means that the group of users didn't specify any concern about the data collection practice being carried by the IoT owner. In this case, no data filter container is launched and the data publisher container reads the information to report to the cloud provider directly from the persistent data store. 
\item  \textbf{Privacy-with data update:} In this situation, a privacy negotiation takes place and the group preference requires that some private information to be filtered from the sensor data before reporting to the cloud. In particular, we use face filtering for captured images in the experiment as an example of a user requesting a privacy measure. The face filter detect all faces in the picture and blur their corresponding pixels.
\item \textbf{Privacy-no data update:} In this situation, a privacy negotiation will take place where the privacy preferences of the group is merged together and considered by the IoT owner. However, contrary to the previous situation, the outcome of the negotiation doesn't require any update for the data.
\end{enumerate}

We report the results for these three scenarios for our choice of data filter (i.e. face detection and blur) in Section \ref{sec:evaluation}. Note that the overhead of the data filter is dependent on the application or sensory interface at hand. Therefore, it is the responsibility of the designer or solution architect to ensure that any kind of applied privacy filters doesn't negatively impact the user experience of the whole IoT application or service. We report in our results the time efficiency for the face detection and blur to demonstrate by an example how a particular application such as the surveillance application could be affected by one possible corresponding privacy filter which is the face detection and blur.

\section{Evaluation}
\label{sec:evaluation}
\subsection{User-Managed IoT}
\label{sec:eval-person}
\begin{figure*}[!t] 
\begin{minipage}[t]{0.32\textwidth}
\includegraphics[width=\textwidth]{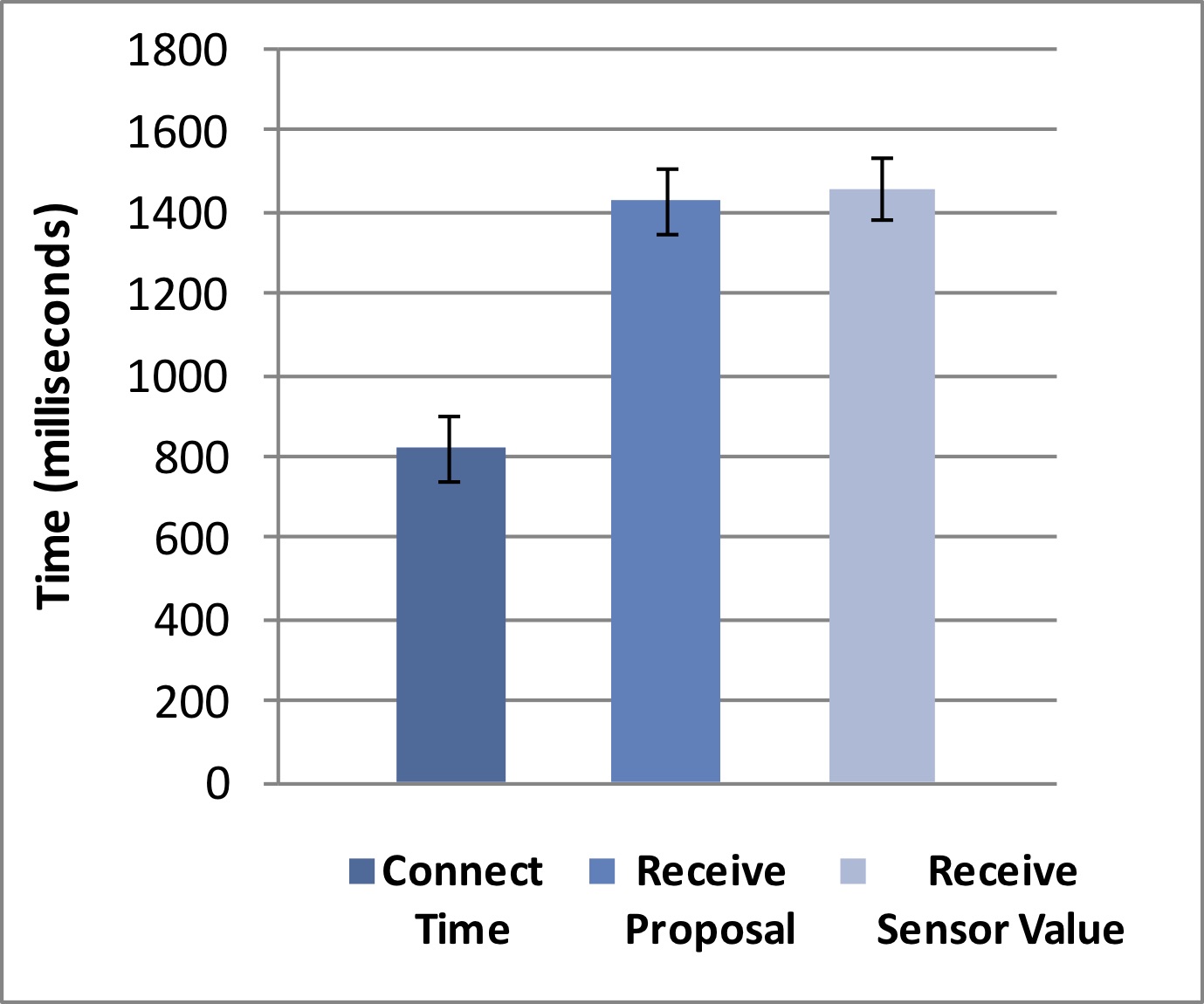} 
\caption{Aggregate time for major negotiation milestones (IoT User\textbackslash User-Managed IoT.)}
\label{fig:phases-per} 
\end{minipage} \hfill
\begin{minipage}[t]{0.32\textwidth}
\includegraphics[width=\textwidth]{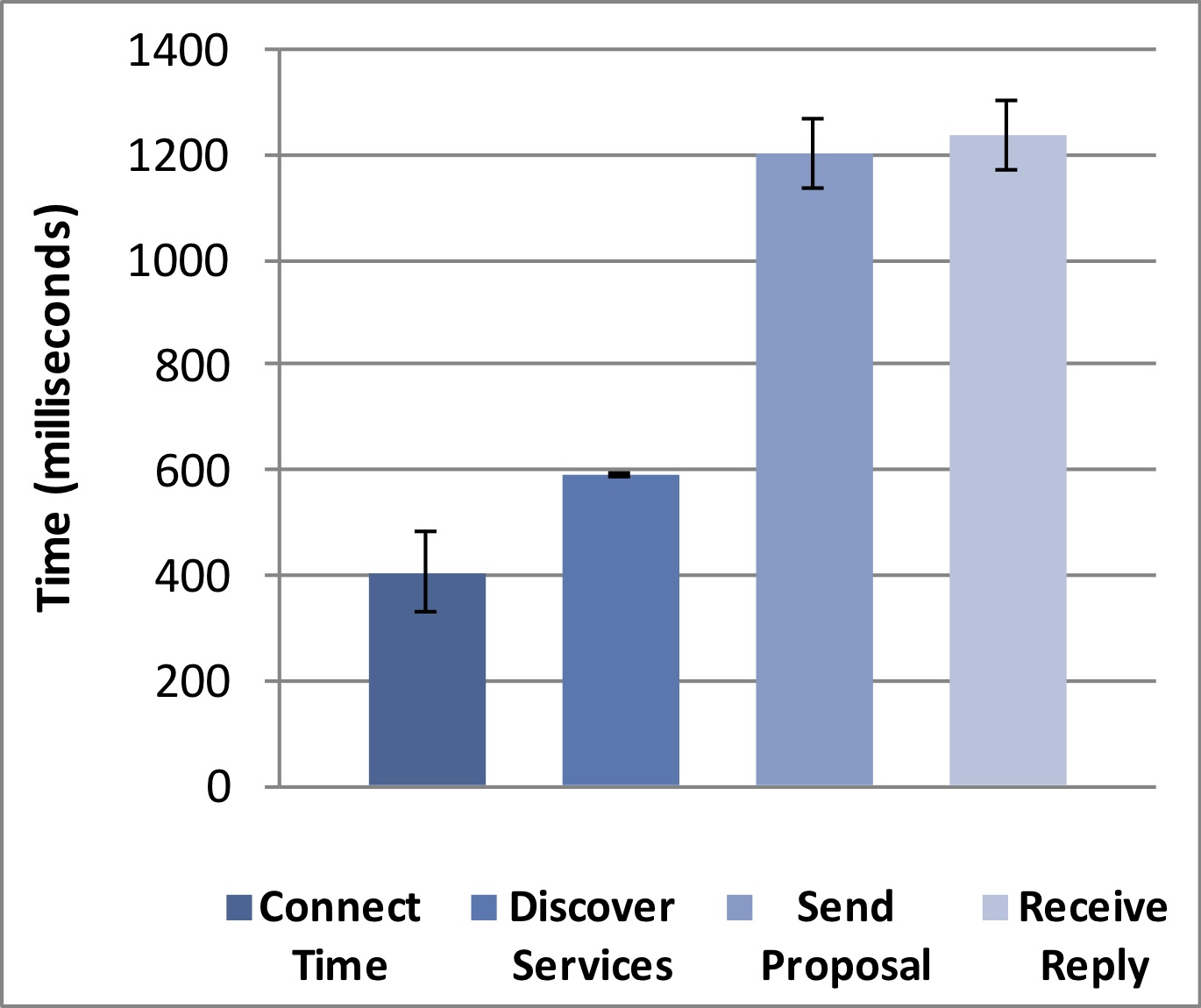} 
\caption{Aggregate time for major negotiation milestones (IoT Owner\textbackslash User-Managed IoT.)}
\label{fig:phases-cen} 
\end{minipage} \hfill
\begin{minipage}[t]{0.32\textwidth}
\includegraphics[width=\textwidth]{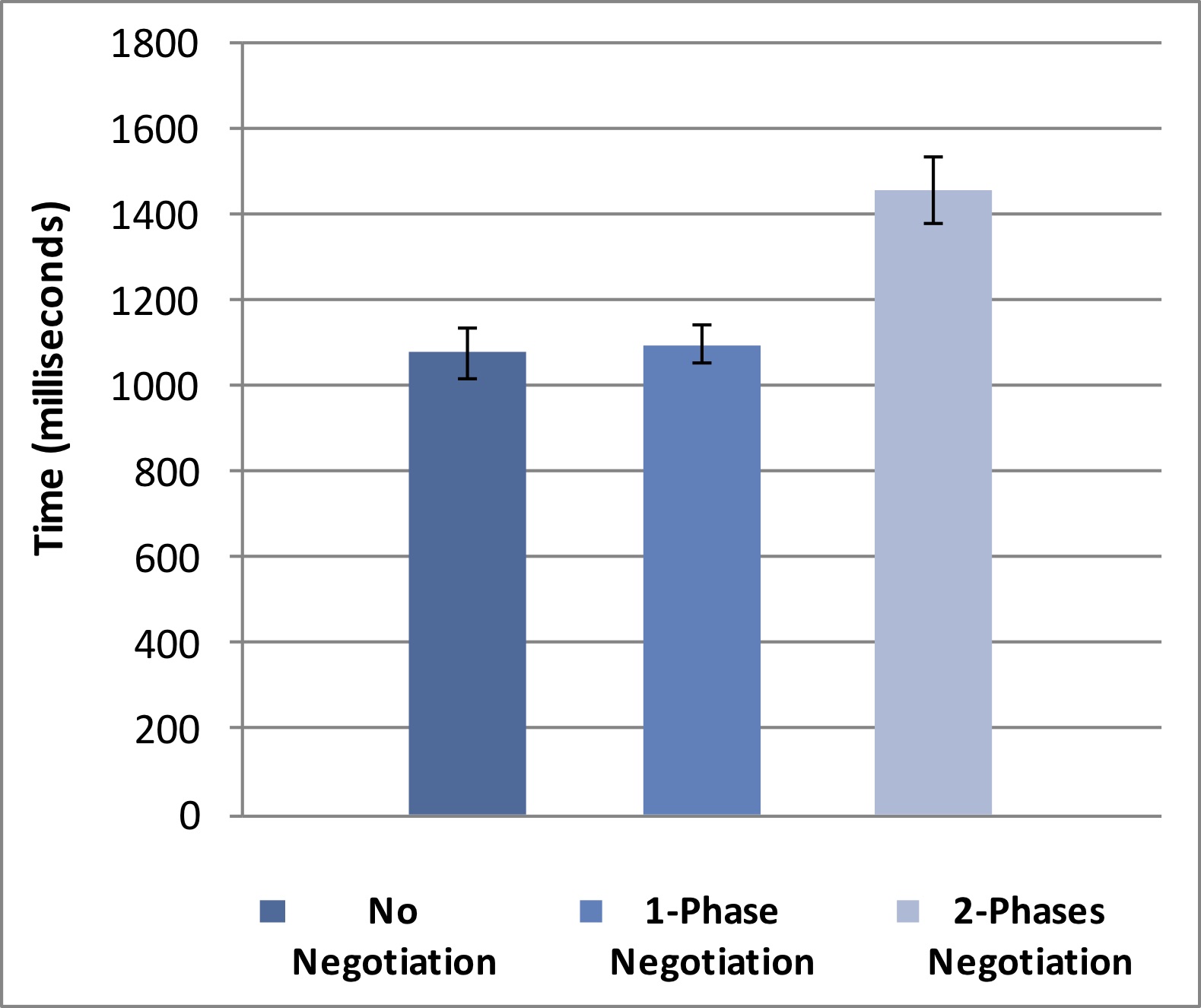} 
\caption{Total time to receive sensor reading for no-negotiation, 1-phase \& 2-phases negotiation in user-managed IoT.}
\label{fig:total-per} 
\end{minipage} \hfill
\end{figure*}
The negotiation protocol in user-managed IoT utilizes BLE to communicate and serve requests for data sharing. A star network is formed whenever an acceptable request arrives from a IoT user in which the IoT owner becomes the central node and the IoT user and any future IoT users will become host nodes. The complexity and time\textbackslash energy efficiency of this process are reported in our previous work 
\citep{alanezi2017leveraging}. 
Note that the IoT owner can serve simultaneous IoT users by simply joining new users to this BLE star network. Also note that we choose to implement this communication mechanism using BLE as it is becoming a standard communication protocol for IoT devices. Nevertheless, the negotiation protocol can be implemented over other communication standards such as ZigBee or NFC. 
Figure \ref{fig:phases-per} shows the aggregate time taken for every phase in the negotiation. The aggregate time is calculated from the beginning of the request (i.e. sending BLE broadcast embedded with data request information) and involves the time taken for the previous phases. We performed each experiment five times and report the average with the standard error on each bar. The figure shows that, after broadcasting a request in a BLE beacon, it takes around 800 ms to be connected to an IoT owner in the place who is willing to support this request. Recall from the design section that after establishing a connection two situations might occur. First, The IoT owner accepts the utility of the request and sends the required sensor data in what we call 1-phase negotiation. The total time for this situation is reported in Figure \ref{fig:total-per} along with the total time for the other two scenarios of no-privacy and 2-phases negotiation. Second, the IoT owner might offer an alternative proposal that is suitable for them. The total time for the IoT user to receive the alternative proposal is 1400 ms on average. We note here that the additional time of ~600 ms is dominated by the time required by the BLE to interrogate the IoT user's device after connecting to it to be able to call its services. Assuming that the proposal is accepted, replying to the proposal and receiving the sensor value requires an additional 30 ms only. Figure \ref{fig:phases-cen} reports aggregate time for the negotiation protocol from the IoT owner's perspective. The average total time to connect to a device after detecting a supported request in a BLE broadcast is 400 ms. This is followed by an almost 600 ms for discovering services on the IoT user's device. We report the connect time and the service discovery separately. The time taken from the moment a beacon is detected to sending an alternative proposal is around 1200 ms on on average. It takes an additional 30 ms on average for the IoT owner's device to receive a reply to their proposal. Finally, we report in Figure \ref{fig:total-per} the average overall time to receive the sensor data for 1-phase communication and 2-phases communication. This time is measured from the moment the IoT user sends a beacon with embedded request information to the moment they actually receives the required sensor data (i.e. the temperature sensor reading). We also include a no-negotiation scenario in which the sensor data is sent to the requester immediately without matching the specification of the request with the privacy requirements of the IoT owner. As expected, the results show that the 1-phase negotiation adds negligible time overhead compared to the negotiation scenario as it only adds information to the beacon and process them at the IoT owner's side. On the other hand, the 2-phases negotiation adds on average 350 ms, which includes the time required to receive an alternative proposal from the IoT owner, reply by accepting the proposal, then receiving the sensor value. 
\begin{figure*}[!t] 
\begin{minipage}[t]{0.32\textwidth}
\includegraphics[width=\textwidth]{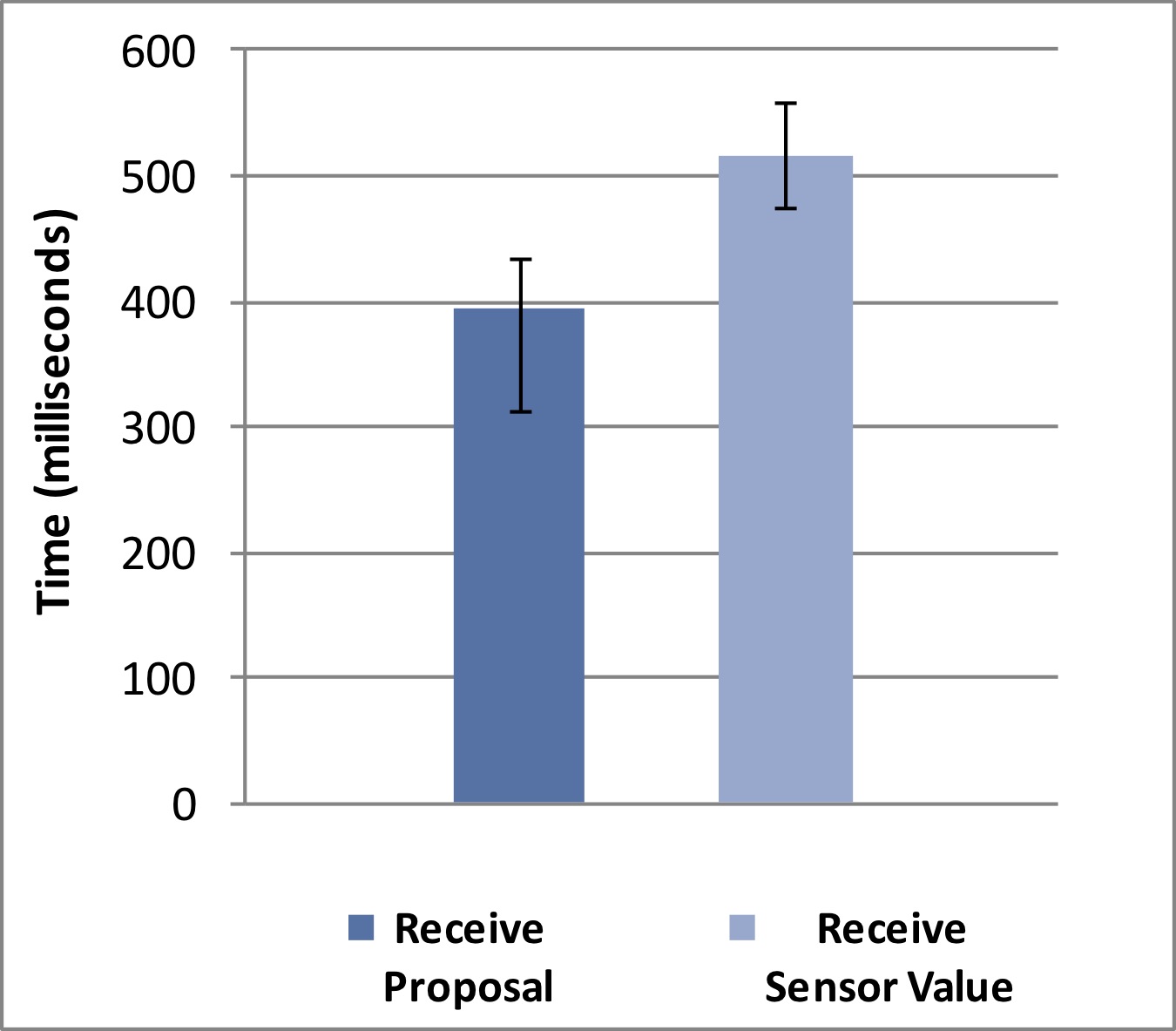} 
\caption{Aggregate time for major negotiation milestones (IoT User\textbackslash Edge-Managed IoT.)}
\label{fig:phases-requester} 
\end{minipage} \hfill
\begin{minipage}[t]{0.32\textwidth}
\includegraphics[width=\textwidth]{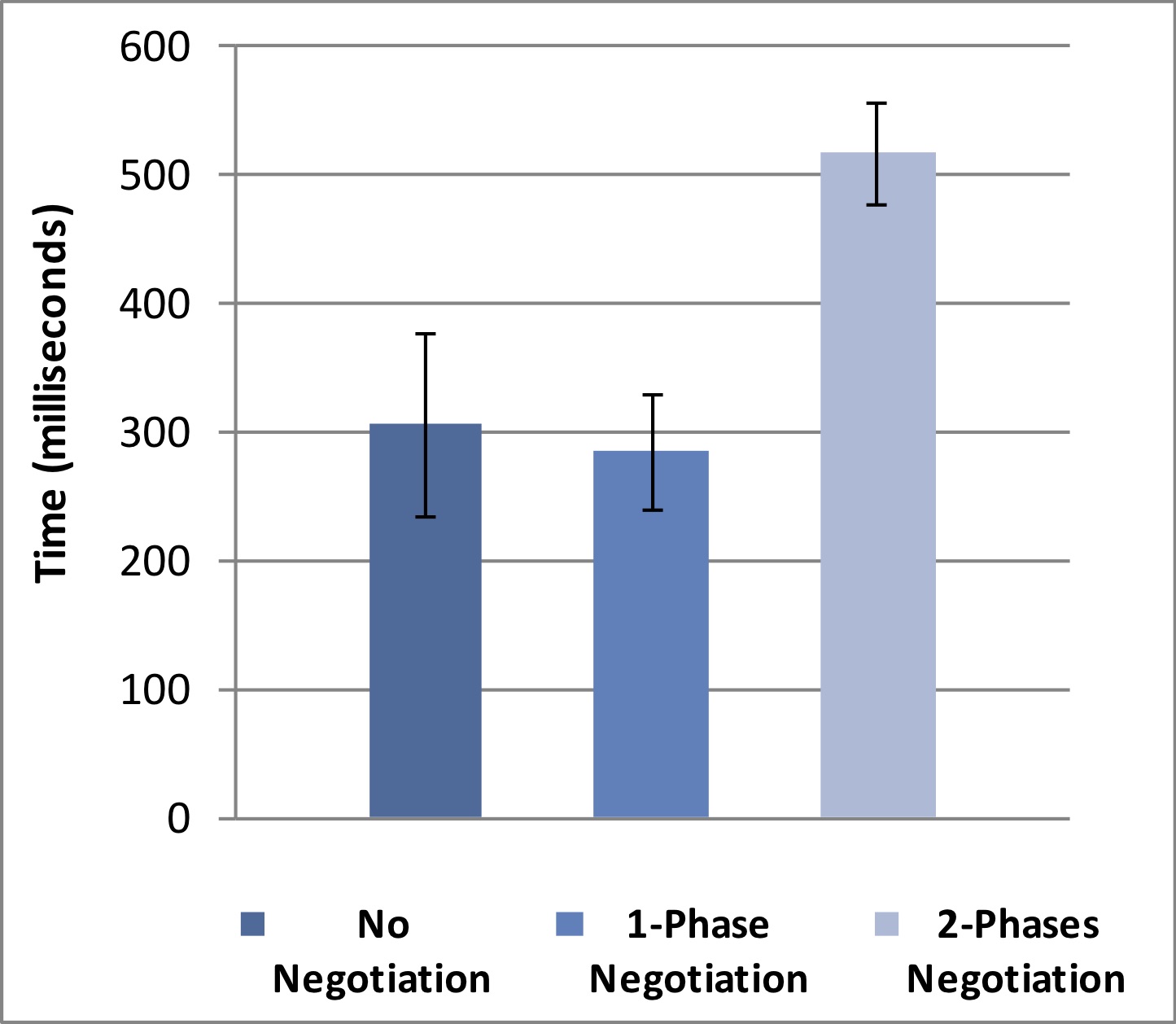} 
\caption{Total time to receive sensor reading for no-negotiation, 1-phase \& 2-phases negotiation in edge-managed IoT.}
\label{fig:strategy-privacy-edge} 
\end{minipage} \hfill
\begin{minipage}[t]{0.32\textwidth}
\includegraphics[width=\textwidth]{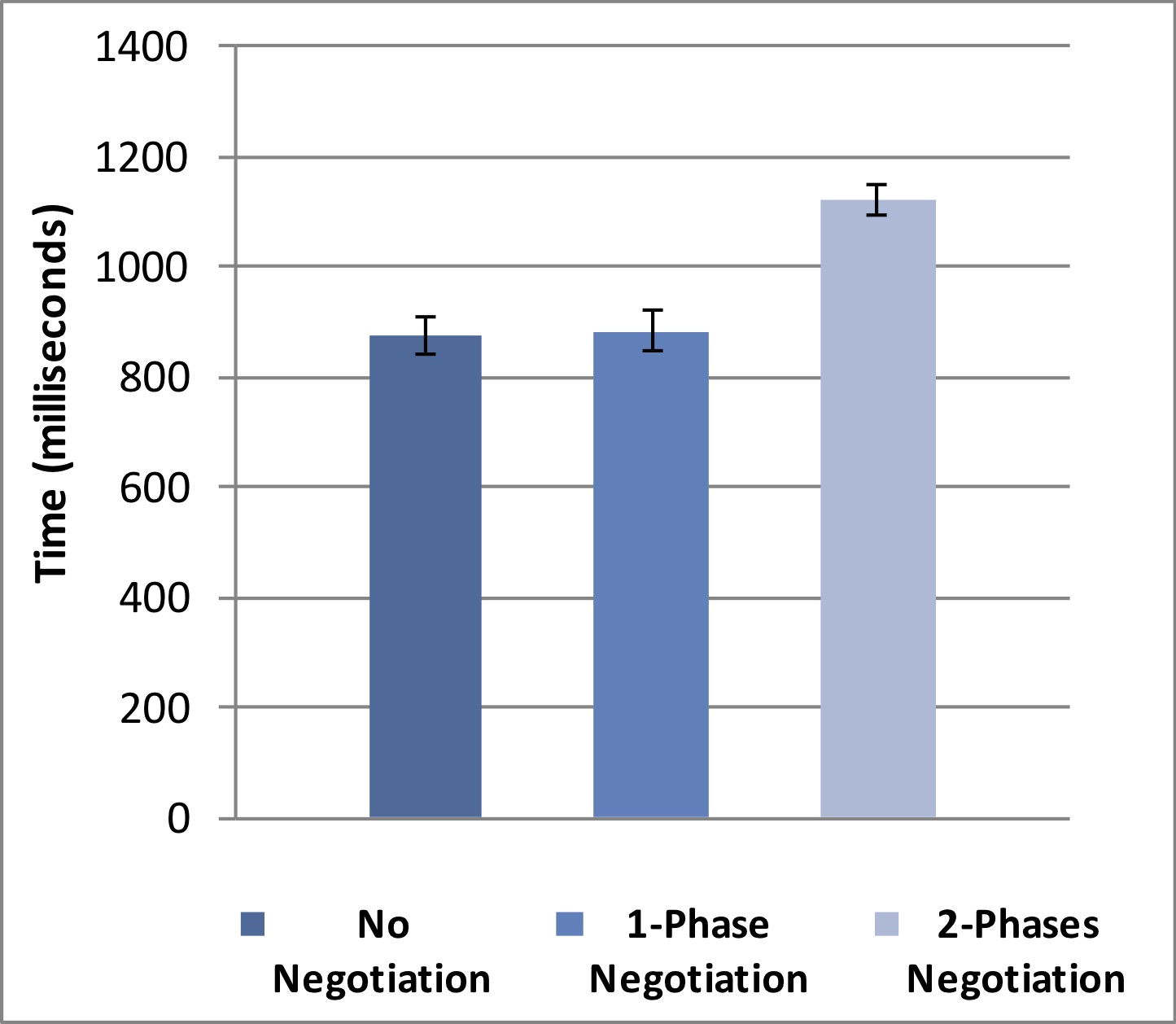} 
\caption{Total time to receive sensor reading for no-negotiation, 1-phase \& 2-phases negotiation in cloud-managed IoT.}
\label{fig:strategy-privacy-cloud} 
\end{minipage} \hfill
\end{figure*}

\subsection{Edge-Managed and Cloud-Managed IoT}
\label{sec:eval-edge}
In this subsection, we repeat the experiments in the previous subsection but with the IoT now being managed by either an edge or a cloud server. First, we begin by describing the results of the edge-managed scenario. The IoT user needs to negotiate with and access the IoT infrastructure through the edge server and all communication is happening over wireless LAN network. Figure \ref{fig:phases-requester} shows the major milestones from the IoT user's side in this negotiation situation. We note two things in this experiment. First, as opposed to the user-managed negotiation scenario, there is no connect phase since the IoT owner is now a server that continuously listens to a network port to reply to requests from IoT users. Hence, there is no connect phase per se. Second, unlike the user-managed IoT experiment, we don't report in this experiment results from the IoT owner's side since the owner is now a server with presumably abundant resources. We see from the figure that the time to receive an alternative proposal from the IoT owner is now less than 400 ms. This time includes the time to send a request to the edge server with an unacceptable privacy settings and receiving back an alternative proposal. Furthermore, the time to receive the sensor value if the IoT user accepted the IoT owner's proposal is around 500 ms. We conclude that managing the IoT infrastructure using an edge server provided much better time efficiency than the user-managed IoT scenario using BLE. We also report in Figure \ref{fig:strategy-privacy-edge} a comparison between three situations for negotiating privacy requirements. There are two important lessons that can be learned from this figure. First, when comparing the no-negotiation scenario with the 1-phase negotiation scenario, we see a difference of around 20 ms. This difference is negligible as it is attributed to the variability of the performance of the wireless LAN, which can be realized from the standard errors. Therefore, similar to the results from the user-managed IoT, these two situations have similar time performance. Second, we see that adding a second negotiation phase added an average of 200 ms attributed to sending the proposal from the IoT owner and receiving a reply back before sending the required sensor reading. Overall, the performance is still better than the same scenario for the user-managed IoT. Finally, Figure \ref{fig:strategy-privacy-cloud} reports similar results for the previous figure but with a cloud-based server now used to manage IoT as opposed to an edge-based server. We notice from the figure that the latency doubled in all three scenarios due the need of routing the communication through the Internet. This result is in harmony with other research \citep{satyanarayanan2017emergence} that highlighted the benefits that cloudlets proximity brings for better management of IoT infrastructure but highlighted various challenges for this paradigm to become a reality.


\subsection{Camera Sensor Group Privacy Scenario Evaluation}
\label{subsec:camera-eval}
The evaluation of the group-based scenario focuses on the effect of various privacy negotiation outcomes on the overall performance. For example, if the group decides to apply some filter to amend the sensory information before reporting to the cloud, this decision incurs an overhead of activating the component responsible for the negotiation as well as the overhead for applying the filter. We measure in our experiment three possible scenarios. First, a baseline situation of \textbf{no-privacy} where the collection of images from the camera sensor is acceptable by definition and there is no privacy negotiation option given to the users. This scenario might correspond to a situation where the IoT owner enforces full image collection due to security reasons or to public places where capturing and publishing of full images is not considered a breach of user privacy (e.g. cameras in public streets). Second, a \textbf{privacy-no-data-update} in which a negotiation had taken place but the result was to report the camera images to the cloud as is without applying a privacy filter. Third, a \textbf{privacy-with-data-update} scenario where the results of the negotiation requires applying face detection and filtering before reporting the images to the cloud.

\begin{figure*}[!t] 
\begin{minipage}[t]{0.32\textwidth}
\includegraphics[width=\textwidth]{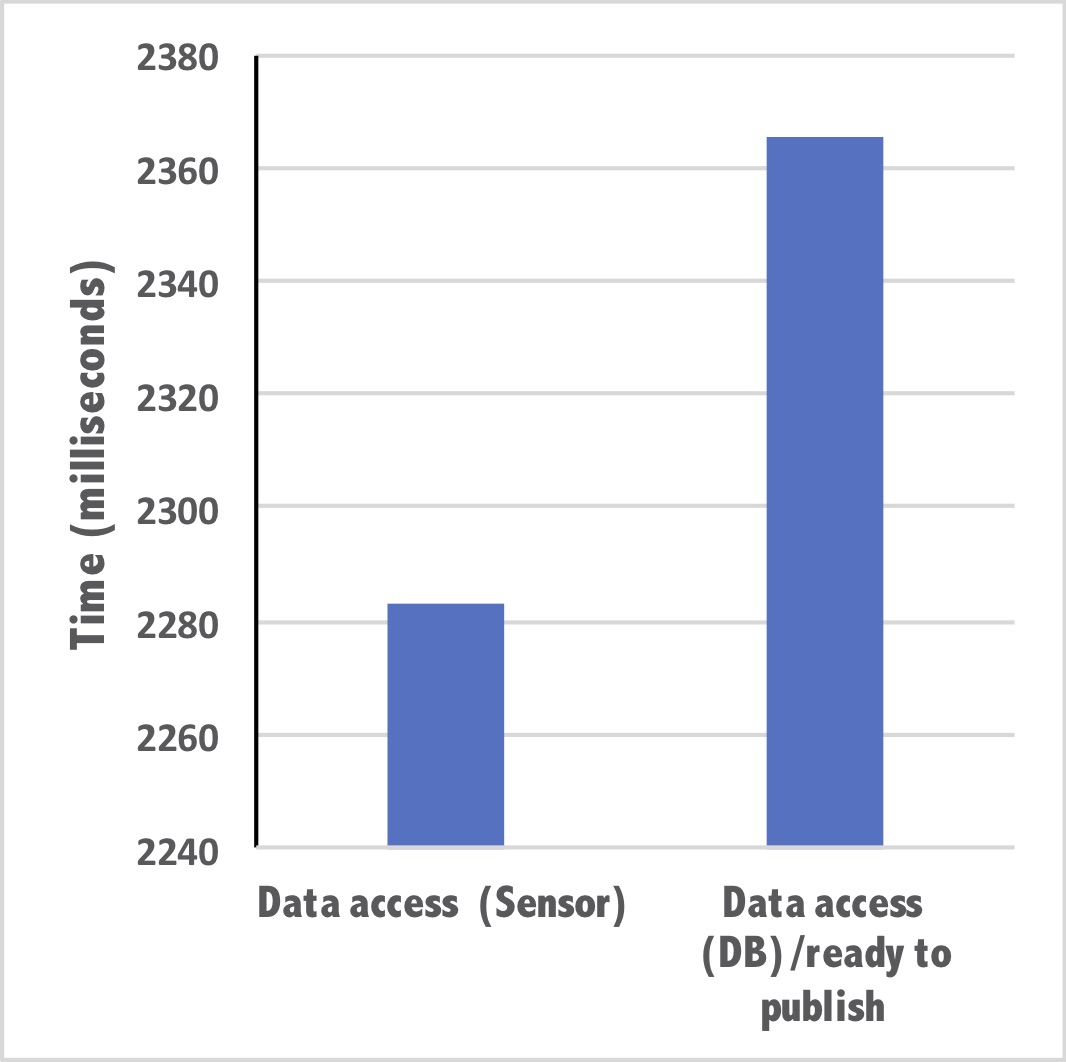} 
\caption{Aggregate time for major milestones in the baseline situation of no privacy (Edge-managed\textbackslash Privacy).}
\label{fig:no-privacy} 
\end{minipage} \hfill
\begin{minipage}[t]{0.32\textwidth}
\includegraphics[width=\textwidth]{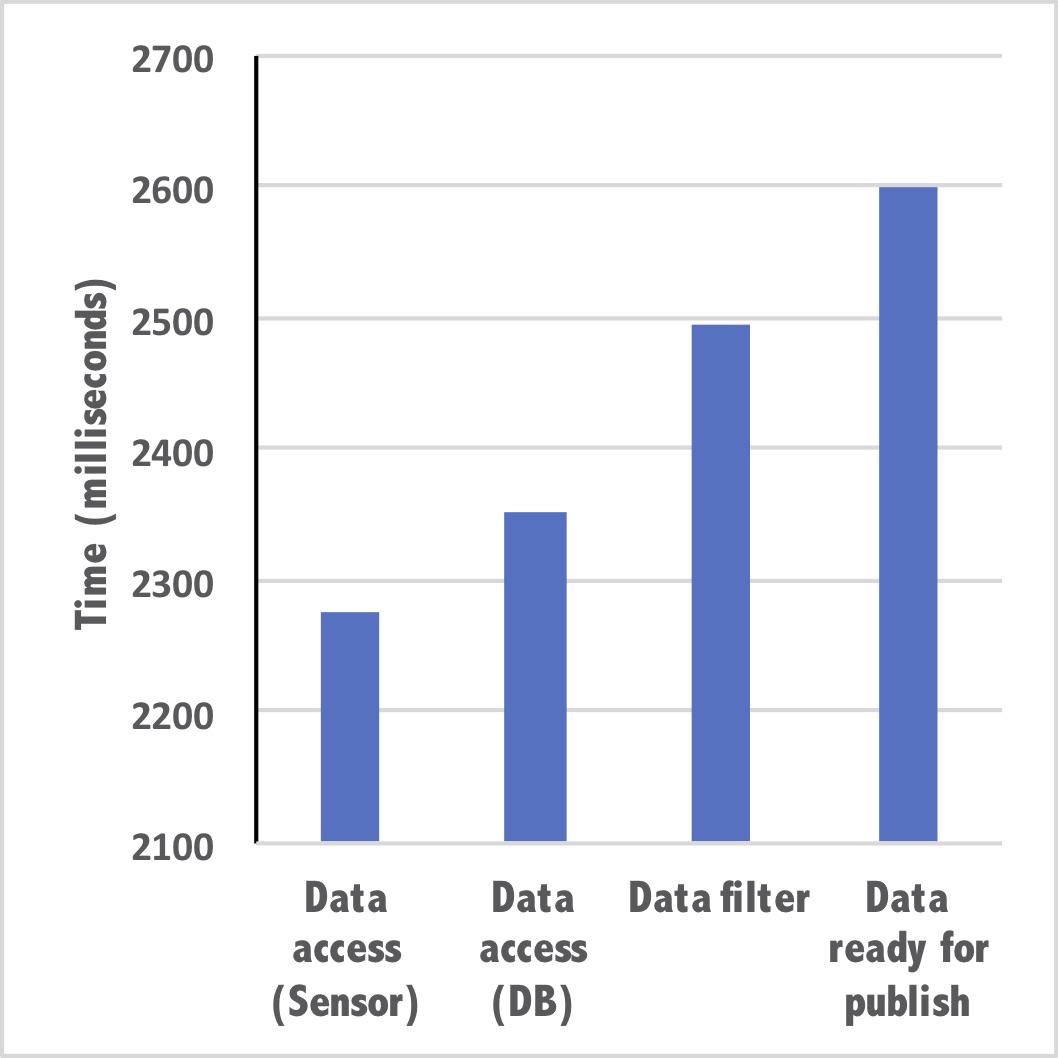} 
\caption{Aggregate time for major milestones in the situation of privacy-without-data-update (Edge-managed\textbackslash Privacy).}
\label{fig:privacy-no-data-update} 
\end{minipage} \hfill
\begin{minipage}[t]{0.32\textwidth}
\includegraphics[width=\textwidth]{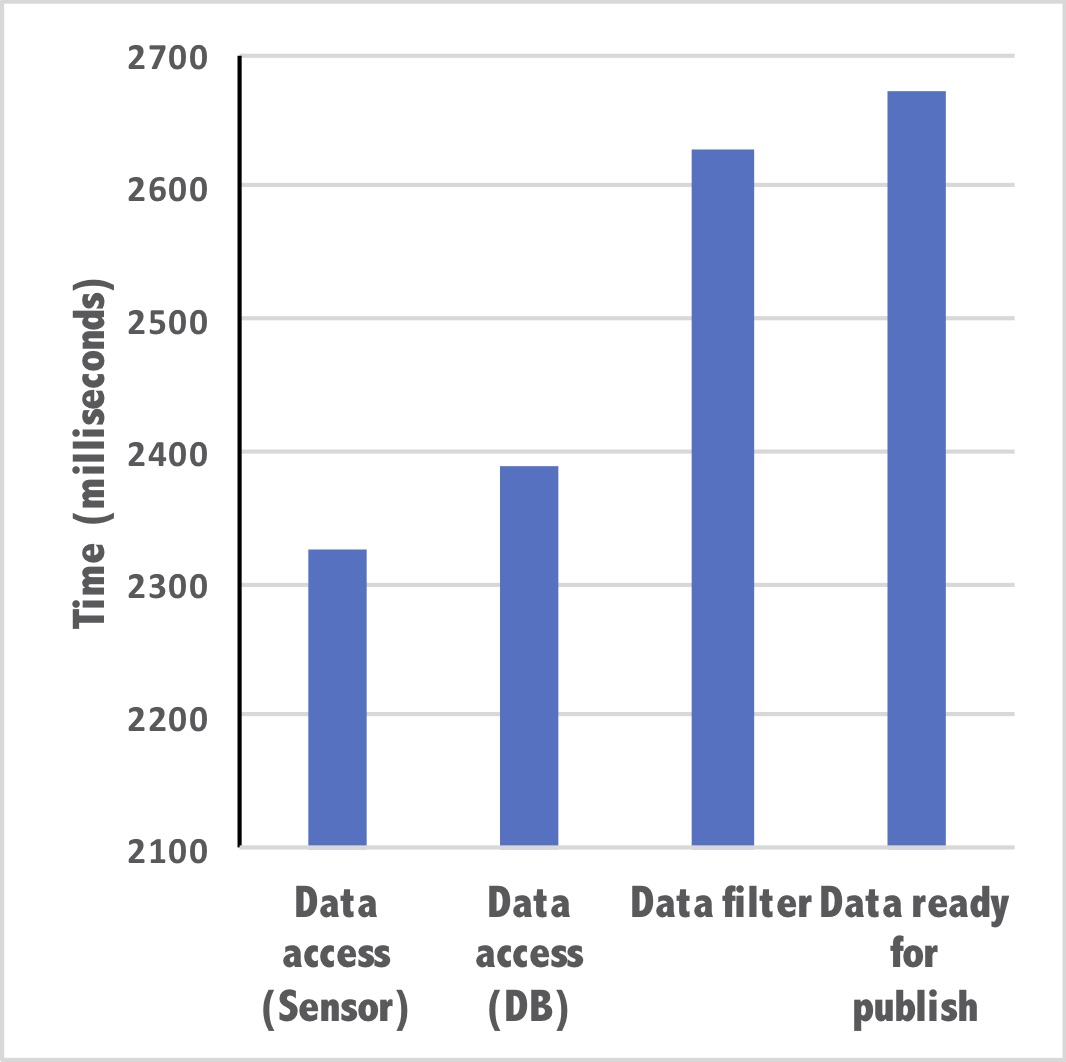} 
\caption{Aggregate time for major milestones in the situation of privacy-with-data-update (Edge-managed\textbackslash Privacy).}
\label{fig:privacy-with-data-update}
\end{minipage} \hfill
\end{figure*}

\begin{figure}
  \includegraphics[width=0.45\textwidth]{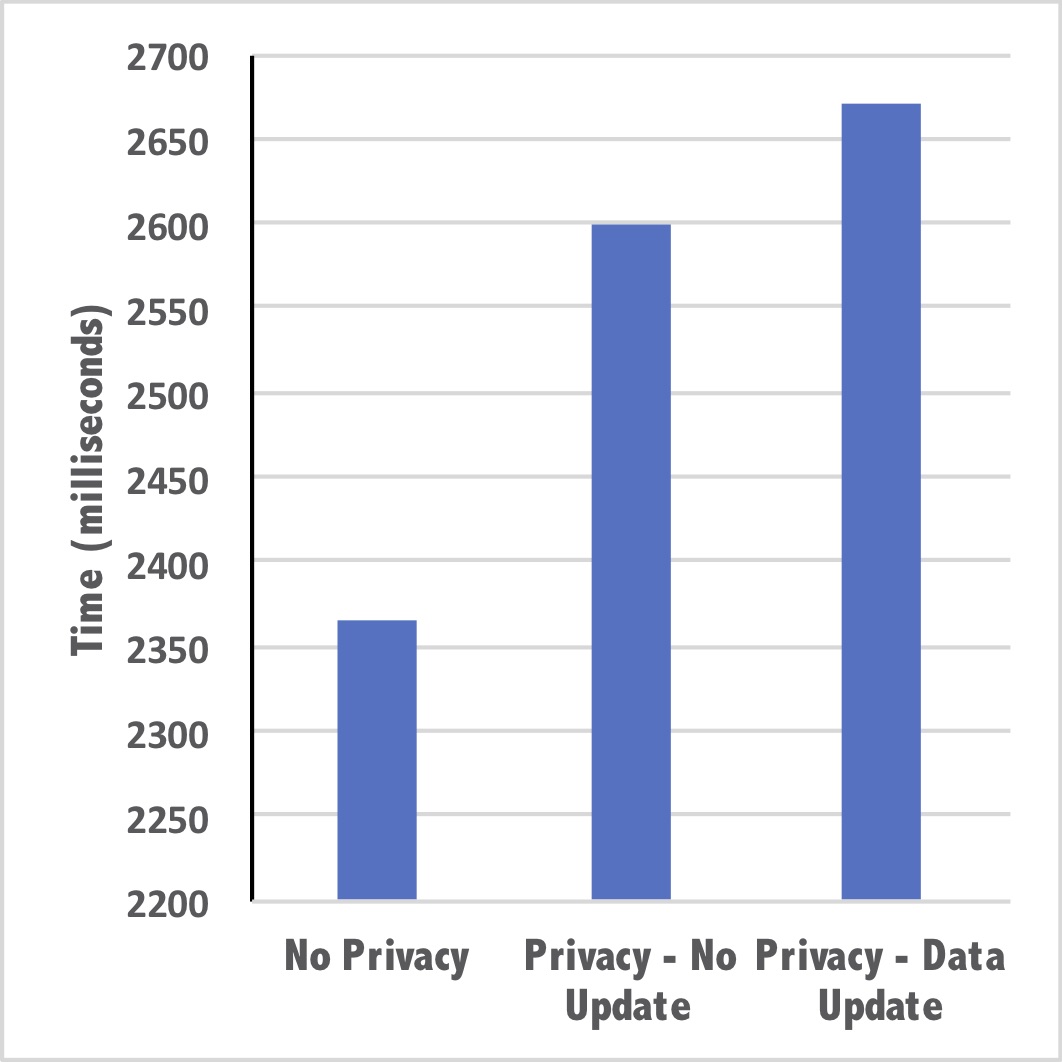}
  \caption{Total time for three situations of no-privacy privacy-no update \& privacy-data update (Edge-managed\textbackslash Privacy).}
  \label{fig:overall}
\end{figure}

Figure \ref{fig:no-privacy} shows the performance for the two milestones required when no privacy is activated in the pipeline architecture. First, we see that the time needed to read the image form the camera sensor by the edge server over Wi-Fi is 2.2 seconds. Afterwards, passing the image over the message queue to data store component to save it to the persistent data store and reading it by the data publisher for reporting to the cloud requires an addition 100 milliseconds. Hence, the overall time required for preparing the data for reporting to the cloud is 2.3 seconds. The experiment in Figure \ref{fig:privacy-no-data-update} reports the time performance for the different milestones when privacy negotiation is activated. Note here that this situation requires full blown activation of all components of the pipeline architecture. Hence, the number of milestones is four in this situation which is double the number of milestones in the previous figure. We see from the figure that the first two milestones have taken similar time to the same two milestones in the previous experiment, which is logical since these are the same steps. We also note that activating the data filter component required reading the data from the persistent data store and passing it to the data filter component to perform the negotiation thereby adding an additional 145 milliseconds time overhead to the pipeline. Finally, an extra 100 milliseconds is required to pass the data from the data filter to data publisher via the message queue dedicated for communication between the two components. When comparing the previous two situations we see that the activation of the data filter component to perform the privacy filtering incurred an additional time overhead of 250 milliseconds. We turn into Figure \ref{fig:privacy-with-data-update} which reports the time performance for a situation similar to the previous figure with one difference where faces are detected in the image and face blur must be applied. We see from the figure that data filter component has taken 240 milliseconds compared to 145 milliseconds in the previous figure. The additional 100 milliseconds can be attributed to the time need to perform the privacy filtering, which is the face blur code in our specific scenario.
We learn from this results that privacy filtering can be costly in terms of time overhead and hence must be planned carefully when introduced to any application. Finally, Figure \ref{fig:overall}, compares the overall timing for all three scenarios. When compared to the baseline scenario of No Privacy, we see that privacy negotiation and its application incurs some overhead, about 10\% in Privacy-No Update scenario and 13\% in Privacy - Data Update scenario.

\section{Discussion}
\label{sec:discussion}
This paper promotes a solution to cover key issues related to system architecture for supporting privacy in the IoT domain. However, the privacy concern in IoT systems is a complex research topic and other open research problems remain at large. In this section, we point out these research challenges and point the reader to possible future directions. 

First, learning the privacy preferences of users is a challenging process. User privacy depends on various factors including the type of the data being collected, the usage scenario and location context (i.e. public vs. private spaces). Fortunately, studies have shown that the privacy preferences can be predicted by observing the user behavior towards privacy using limited number of scenarios \citep{naeini2017privacy}. Also, social cues can be given to users to allow them to make informed decisions about their privacy decisions in an IoT environment \citep{emami2018influence}. Despite the great benefits these studies provide, a missing piece of the puzzle is to observe the user behavior in a real IoT environment, as these studies depend largely on user questionnaires. By learning from studies from a real world IoT platform, solutions like our framework can adapt techniques for learning users privacy preferences without depending on cumbersome preferences collection screens. The learned privacy preference that can resemble a trained prediction model can be pushed from the cloud, where the model training happened, to users' mobile devices, where inference for privacy preferences can happen based on the input from the environment, such as the current location or social context of the user and the data collection requirements.

Another important IoT challenge that is also applicable to our framework is scalability. An IoT environment can involve large number of devices and it is necessary that the preferences of users are negotiated and applied to these devices in a timely manner. Our framework uses BLE as the communication medium between the IoT owner, IoT users and IoT devices in the user-managed scenario. Besides its great energy efficiency, BLE is under continuous improvements with BLE 5 \citep{ble-5} providing double the bandwidth and 4-fold increase in communication range compared to the previous BLE version. The framework also utilizes Wi-Fi, which has also seen recent improvements by introducing the IEEE 802.11ah standard \citep{wifi-ah} aimed at addressing issues relevant to the IoT, such as energy efficiency and scalability.

Finally, a challenge that also must be considered is the added overhead for applying privacy filters. IoT sensors data filtering mechanisms are dependent on the type of sensor being filtered, and their associated cost can vary from adding negligible to adding substantial time or energy overhead. For example, detecting and removing particular sounds or specific faces from a video feed requires running pre-trained models, which can be costly, thereby negatively impacting the quality of the service. Hence, it is important that system designers choose the right filtering mechanism that preserves the quality of the service while achieving the needed privacy for user data. There are various research efforts aimed at efficiently introducing privacy measures to collected data from an IoT environment without impacting the utility of the service provided. For example, an edge-based infrastructure is proposed in an IoT environment \citep{das2017assisting} to detect and denature faces from a video feed based on user privacy preference. In addition, privacy-aware offloading is used in healthcare IoT environments \citep{min2018learning} to protect the user location and usage patterns while utilizing resources from the mobile-edge.

\section{Conclusion}
\label{sec:conclusion}
This paper presented a privacy negotiation scheme to address the privacy requirements of users in IoT environments. The proposed approach is practical as it negotiates the privacy policy of the user with the IoT owner without user intervention and supports the selection from among multiple predefined IoT user and owner privacy policies. The paper addresses privacy negotiation in both individual and group contexts. The feasibility of the negotiation protocol was demonstrated by means of a thorough implementation and evaluation over three widely accepted IoT scenarios.


\bibliography{refs2}   

\begin{thebibliography}{31}
\providecommand{\natexlab}[1]{#1}
\providecommand{\url}[1]{\texttt{#1}}
\expandafter\ifx\csname urlstyle\endcsname\relax
  \providecommand{\doi}[1]{doi: #1}\else
  \providecommand{\doi}{doi: \begingroup \urlstyle{rm}\Url}\fi

\bibitem[Alanezi and Mishra(2018)]{alanezi2018privacy}
Alanezi, K. and Mishra, S.
\newblock A privacy negotiation mechanism for the internet of things.
\newblock In \emph{2018 IEEE 16th Intl Conf on Dependable, Autonomic and Secure
  Computing, 16th Intl Conf on Pervasive Intelligence and Computing, 4th Intl
  Conf on Big Data Intelligence and Computing and Cyber Science and Technology
  Congress (DASC/PiCom/DataCom/CyberSciTech)}, pages 512--519. IEEE, 2018.

\bibitem[Alanezi et~al.(2017)Alanezi, Rafiq, Chen, and
  Mishra]{alanezi2017leveraging}
Alanezi, K., Rafiq, R.~I., Chen, L., and Mishra, S.
\newblock Leveraging ble and social trust to enable mobile in situ
  collaborations.
\newblock In \emph{Proceedings of the 11th International Conference on
  Ubiquitous Information Management and Communication}, page~98. ACM, 2017.

\bibitem[{{Bluetooth SIG}}(2016)]{ble-5}
{{Bluetooth SIG}}.
\newblock {Bluetooth Core Specification 5.0 FAQ}, 2016.
\newblock URL \url{https://www.mouser.com/pdfdocs/bluetooth-5-faq.pdf}.
\newblock [Accessed Dec 27th, 2020].

\bibitem[Broenink et~al.(2010)Broenink, Hoepman, Hof, Van~Kranenburg, Smits,
  and Wisman]{broenink2010privacy}
Broenink, G., Hoepman, J.-H., Hof, C.~v., Van~Kranenburg, R., Smits, D., and
  Wisman, T.
\newblock The privacy coach: Supporting customer privacy in the internet of
  things.
\newblock \emph{arXiv preprint arXiv:1001.4459}, 2010.

\bibitem[Cranor(2002)]{cranor2002web}
Cranor, L.
\newblock \emph{Web privacy with P3P}.
\newblock " O'Reilly Media, Inc.", 2002.

\bibitem[Das et~al.(2017)Das, Degeling, Wang, Wang, Sadeh, and
  Satyanarayanan]{das2017assisting}
Das, A., Degeling, M., Wang, X., Wang, J., Sadeh, N., and Satyanarayanan, M.
\newblock Assisting users in a world full of cameras: A privacy-aware
  infrastructure for computer vision applications.
\newblock In \emph{2017 IEEE Conference on Computer Vision and Pattern
  Recognition Workshops (CVPRW)}, pages 1387--1396. IEEE, 2017.

\bibitem[Davies et~al.(2016)Davies, Taft, Satyanarayanan, Clinch, and
  Amos]{davies2016privacy}
Davies, N., Taft, N., Satyanarayanan, M., Clinch, S., and Amos, B.
\newblock Privacy mediators: Helping iot cross the chasm.
\newblock In \emph{Proceedings of the 17th International Workshop on Mobile
  Computing Systems and Applications}, pages 39--44. ACM, 2016.

\bibitem[{Docker, Inc.}(2013)]{docker}
{Docker, Inc.}
\newblock Docker containers, 2013.
\newblock URL \url{https://www.docker.com/}.
\newblock [Accessed Dec 27th, 2020].

\bibitem[Dorri et~al.(2017)Dorri, Kanhere, Jurdak, and
  Gauravaram]{dorri2017blockchain}
Dorri, A., Kanhere, S.~S., Jurdak, R., and Gauravaram, P.
\newblock Blockchain for iot security and privacy: The case study of a smart
  home.
\newblock In \emph{2017 IEEE international conference on pervasive computing
  and communications workshops (PerCom workshops)}, pages 618--623. IEEE, 2017.

\bibitem[Dragoni et~al.(2017)Dragoni, Giallorenzo, Lafuente, Mazzara, Montesi,
  Mustafin, and Safina]{dragoni2017microservices}
Dragoni, N., Giallorenzo, S., Lafuente, A.~L., Mazzara, M., Montesi, F.,
  Mustafin, R., and Safina, L.
\newblock Microservices: yesterday, today, and tomorrow.
\newblock In \emph{Present and ulterior software engineering}, pages 195--216.
  Springer, 2017.

\bibitem[Emami~Naeini et~al.(2018)Emami~Naeini, Degeling, Bauer, Chow, Cranor,
  Haghighat, and Patterson]{emami2018influence}
Emami~Naeini, P., Degeling, M., Bauer, L., Chow, R., Cranor, L.~F., Haghighat,
  M.~R., and Patterson, H.
\newblock The influence of friends and experts on privacy decision making in
  iot scenarios.
\newblock \emph{Proceedings of the ACM on Human-Computer Interaction},
  2\penalty0 (CSCW):\penalty0 1--26, 2018.

\bibitem[Fogues et~al.(2017)Fogues, Murukannaiah, Such, and
  Singh]{fogues2017sosharp}
Fogues, R.~L., Murukannaiah, P.~K., Such, J.~M., and Singh, M.~P.
\newblock Sosharp: Recommending sharing policies in multiuser privacy
  scenarios.
\newblock \emph{IEEE Internet Computing}, 21\penalty0 (6):\penalty0 28--36,
  2017.

\bibitem[Henze et~al.(2014)Henze, Hermerschmidt, Kerpen, H{\"a}u{\ss}ling,
  Rumpe, and Wehrle]{henze2014user}
Henze, M., Hermerschmidt, L., Kerpen, D., H{\"a}u{\ss}ling, R., Rumpe, B., and
  Wehrle, K.
\newblock User-driven privacy enforcement for cloud-based services in the
  internet of things.
\newblock In \emph{2014 International Conference on Future Internet of Things
  and Cloud}, pages 191--196. IEEE, 2014.

\bibitem[Hu et~al.(2012)Hu, Ahn, and Jorgensen]{hu2012multiparty}
Hu, H., Ahn, G.-J., and Jorgensen, J.
\newblock Multiparty access control for online social networks: model and
  mechanisms.
\newblock \emph{IEEE Transactions on Knowledge and Data Engineering},
  25\penalty0 (7):\penalty0 1614--1627, 2012.

\bibitem[{IEEE Working Group for WLAN Standards}(2017)]{wifi-ah}
{IEEE Working Group for WLAN Standards}.
\newblock {IEEE Publishes 802.11ah-2016 Standard Amendment}, 2017.
\newblock URL \url{https://standards.ieee.org/standard/802_11ah-2016.html}.
\newblock [Accessed Dec 27th, 2020].

\bibitem[Lampinen et~al.(2011)Lampinen, Lehtinen, Lehmuskallio, and
  Tamminen]{lampinen2011we}
Lampinen, A., Lehtinen, V., Lehmuskallio, A., and Tamminen, S.
\newblock We're in it together: interpersonal management of disclosure in
  social network services.
\newblock In \emph{Proceedings of the SIGCHI conference on human factors in
  computing systems}, pages 3217--3226. ACM, 2011.

\bibitem[Min et~al.(2018)Min, Wan, Xiao, Chen, Xia, Wu, and
  Dai]{min2018learning}
Min, M., Wan, X., Xiao, L., Chen, Y., Xia, M., Wu, D., and Dai, H.
\newblock Learning-based privacy-aware offloading for healthcare iot with
  energy harvesting.
\newblock \emph{IEEE Internet of Things Journal}, 6\penalty0 (3):\penalty0
  4307--4316, 2018.

\bibitem[Mittelstadt(2017)]{mittelstadt2017individual}
Mittelstadt, B.
\newblock From individual to group privacy in big data analytics.
\newblock \emph{Philosophy \& Technology}, 30\penalty0 (4):\penalty0 475--494,
  2017.

\bibitem[Naeini et~al.(2017)Naeini, Bhagavatula, Habib, Degeling, Bauer,
  Cranor, and Sadeh]{naeini2017privacy}
Naeini, P.~E., Bhagavatula, S., Habib, H., Degeling, M., Bauer, L., Cranor,
  L.~F., and Sadeh, N.
\newblock Privacy expectations and preferences in an iot world.
\newblock In \emph{Thirteenth Symposium on Usable Privacy and Security SOUPS
  2017}, pages 399--412, 2017.

\bibitem[{Oracle Corporation}(2012)]{jms}
{Oracle Corporation}.
\newblock Oracle java message service (jms) interface, 2012.
\newblock URL
  \url{https://docs.oracle.com/cd/B19306_01/server.102/b14257/jm_create.htm}.
\newblock [Accessed July 8th, 2019].

\bibitem[Pratama et~al.(2012)Pratama, Hidayat, et~al.]{pratama2012smartphone}
Pratama, A.~R., Hidayat, R., et~al.
\newblock Smartphone-based pedestrian dead reckoning as an indoor positioning
  system.
\newblock In \emph{2012 International Conference on System Engineering and
  Technology (ICSET)}, pages 1--6. IEEE, 2012.

\bibitem[Preibusch(2006)]{preibusch2006implementing}
Preibusch, S.
\newblock Implementing privacy negotiations in e-commerce.
\newblock In \emph{Asia-Pacific Web Conference}, pages 604--615. Springer,
  2006.

\bibitem[Satyanarayanan(2017)]{satyanarayanan2017emergence}
Satyanarayanan, M.
\newblock The emergence of edge computing.
\newblock \emph{Computer}, 50\penalty0 (1):\penalty0 30--39, 2017.

\bibitem[Squicciarini et~al.(2009)Squicciarini, Shehab, and
  Paci]{squicciarini2009collective}
Squicciarini, A.~C., Shehab, M., and Paci, F.
\newblock Collective privacy management in social networks.
\newblock In \emph{Proceedings of the 18th international conference on World
  wide web}, pages 521--530. ACM, 2009.

\bibitem[Stankovic(2014)]{stankovic2014research}
Stankovic, J.~A.
\newblock Research directions for the internet of things.
\newblock \emph{IEEE Internet of Things Journal}, 1\penalty0 (1):\penalty0
  3--9, 2014.

\bibitem[Such and Criado(2016)]{such2016resolving}
Such, J.~M. and Criado, N.
\newblock Resolving multi-party privacy conflicts in social media.
\newblock \emph{IEEE Transactions on Knowledge and Data Engineering},
  28\penalty0 (7):\penalty0 1851--1863, 2016.

\bibitem[Such and Rovatsos(2016)]{such2016privacy}
Such, J.~M. and Rovatsos, M.
\newblock Privacy policy negotiation in social media.
\newblock \emph{ACM Transactions on Autonomous and Adaptive Systems (TAAS)},
  11\penalty0 (1):\penalty0 4, 2016.

\bibitem[{The Eclipse Foundation}(2018)]{kura}
{The Eclipse Foundation}.
\newblock Eclipse kura, 2018.
\newblock URL \url{https://www.eclipse.org/kura/}.
\newblock [Accessed Dec 27th, 2020].

\bibitem[Thomas et~al.(2010)Thomas, Grier, and Nicol]{thomas2010unfriendly}
Thomas, K., Grier, C., and Nicol, D.~M.
\newblock unfriendly: Multi-party privacy risks in social networks.
\newblock In \emph{International Symposium on Privacy Enhancing Technologies
  Symposium}, pages 236--252. Springer, 2010.

\bibitem[Wyatt et~al.(2007)Wyatt, Choudhury, and Bilmes]{wyatt2007conversation}
Wyatt, D., Choudhury, T., and Bilmes, J.
\newblock Conversation detection and speaker segmentation in privacy-sensitive
  situated speech data.
\newblock In \emph{Eighth Annual Conference of the International Speech
  Communication Association}, 2007.

\bibitem[Ziegeldorf et~al.(2014)Ziegeldorf, Morchon, and
  Wehrle]{ziegeldorf2014privacy}
Ziegeldorf, J.~H., Morchon, O.~G., and Wehrle, K.
\newblock Privacy in the internet of things: threats and challenges.
\newblock \emph{Security and Communication Networks}, 7\penalty0 (12):\penalty0
  2728--2742, 2014.

\end{thebibliography}

\end{document}